\documentclass[twocolumn,10pt]{IEEEtran}
\usepackage{stfloats}
\usepackage{xcolor}
\usepackage{amssymb}
\usepackage{theorem,amsbsy,pifont,verbatim,amsfonts,amssymb,color}
\usepackage{xspace,epsfig,graphicx,subfigure,latexsym,fancyhdr,cite}
\usepackage{multirow}
\usepackage{epstopdf}
\usepackage{makecell}
\usepackage{bm}
\usepackage{booktabs}
\usepackage{indentfirst}
\usepackage{cuted}
\usepackage{amsmath}
\usepackage{diagbox}
\usepackage{enumerate}
\usepackage{float}
\usepackage{url}
\usepackage{mdwmath}
\usepackage{mdwtab}
\usepackage{amsmath}
\usepackage{xcolor}
\usepackage{caption}
\usepackage{graphfig}
\usepackage{enumitem}
\usepackage{array}
\usepackage{tabularx}
\usepackage[colorlinks,citecolor=blue,linkcolor=blue,urlcolor=blue]{hyperref}
\usepackage{lipsum}

\def\0{\boldsymbol 0}
\def\1{\boldsymbol 1}

\allowdisplaybreaks

\ifCLASSINFOpdf
\else
\fi

\usepackage{stfloats}

\newcounter{remarkcount}
\newenvironment{remark}[1][]{\refstepcounter{remarkcount}\par\medskip\noindent\textbf{Remark~\theremarkcount~#1.} \itshape}{\par\medskip}

\newcounter{corollarycount}
\newenvironment{corollary}[1][]{\refstepcounter{corollarycount}\par\medskip\noindent\textbf{Corollary~\thecorollarycount~#1.} \itshape}{\par\medskip}

\begin{document}
	%
	\title{\LARGE Near-Field NLOS Localization via Position-Unknown HRIS:\\ From Self-Localization to Target Positioning}
	%
	%
	%
	%

	\author{Hua~Chen,  
		Linke Yu,
		Tuo Wu,  Maged Elkashlan,
		Naofal Al-Dhahir, \emph{Fellow, IEEE},\\ M\'{e}rouane Debbah,  \emph{Fellow},  \emph{IEEE},   and~K. C. Ho,~\IEEEmembership{Fellow,~IEEE}	
		\thanks{{The} work {of Hua Chen, and Linke Yu} was supported by the National Natural Science Foundation of China under Grant 62571286, by the ``Pioneer" and ``Leading Goose" R \& D Program of Zhejiang Province under Grant 2024C01105, and by the China Scholarship Council under Grant 202408330215.  (\emph{Corresponding authors: Tuo Wu}.)}
		\thanks{Hua Chen and Linke Yu are with the Faculty of Electrical Engineering and Computer Science, Ningbo University, Ningbo 315211, China. (E-mail: $\rm dkchenhua0714@hotmail.com$). T. Wu  is with the School of Electronic and Information Engineering, South China University of Technology, Guangzhou 510640, China  (E-mail: $ \rm wtopp0415@163.com$). M. Elkashlan is with the School of Electronic Engineering and Computer Science at Queen Mary University of London, London E1 4NS, U.K. (E-mail: $\rm maged.elkashlan@qmul.ac.uk$).  Naofal Al-Dhahir is with the Department of Electrical and Computer Engineering, The University of Texas at Dallas, Richardson, TX 75080 USA (E-mail: $\rm aldhahir@utdallas.edu$). M. Debbah is with  KU 6G Research Center, Department of Computer and Information Engineering, Khalifa University, Abu Dhabi 127788, UAE (E-mail: $\rm merouane.debbah@ku.ac.ae$). K. C. Ho is with the Department of Electrical Engineering and Computer Science, University of Missouri, Columbia, MO 65211, USA (E-mail: $\rm hod@missouri.edu$). }
	}

	\markboth{IEEE TRANSACTIONS ON WIRELESS COMMUNICATIONS~Vol.~XX, No.~XX, XX~2026}%
	{Shell \MakeLowercase{\textit{et al.}}: A Sample Article Using IEEEtran.cls for IEEE Journals}
	\maketitle
	
	\begin{abstract}
		Current reconfigurable intelligent surface (RIS)-aided near-field (NF) localization methods assume the RIS position is known \textit{a priori}{, and it} has limited their practical applicability. This paper applies a hybrid RIS (HRIS) at an unknown position to locate non-line-of-sight (NLOS) NF targets.  To this end, we first propose a two-stage gridless localization framework for achieving HRIS self-localization, and then determine the positions of the NF targets. In the first stage, we use the NF Fresnel approximation to convert the signal model into a virtual far-field model through delay-based cross-correlation of centrally symmetric HRIS elements.  Such a conversion will naturally extend the aperture of the virtual array. A single-snapshot decoupled atomic norm minimization (DANM) algorithm is then proposed to locate an NF target relative to the HRIS, which includes a two-dimensional (2-D) direction of arrival (DOA) estimation with automatic pairing, the multiple signal classification (MUSIC) method for range estimation, and a total least squares (TLS) method to eliminate the Fresnel approximation error. In the second stage, we leverage the unique capability of HRIS in simultaneous sensing and reflection to estimate the HRIS-to-base station (BS) direction vectors using atomic norm minimization (ANM), and derive the three-dimensional (3-D) HRIS position with two BSs via the least squares (LS)-based geometric triangulation. Furthermore, we propose a semidefinite relaxation (SDR)-based HRIS phase optimization method to enhance the received signal power at the BSs, thereby improving the HRIS localization accuracy, which, in turn, enhances NF target positionings. The Cram\'er-Rao bound (CRB) for the NF target parameters and the position error bound (PEB) for the HRIS coordinates are derived as performance benchmarks. Simulation results validate the effectiveness of the proposed framework in localizing NLOS NF targets without prior knowledge of the HRIS position.
	\end{abstract}
	

	%
	\IEEEpeerreviewmaketitle

	\section{Introduction}
	%
	%
	%
	
	In recent years, breakthroughs in wireless networking capabilities have made positioning technology a prominent research focus. The next sixth-generation (6G) networks will leverage millimeter-wave frequencies and ultra-wide bandwidth to achieve super-resolution positioning \cite{2,3,Tuo2,4}, among which high-precision direction of arrival (DOA) estimation has been regarded as one of the key technologies. However, non-line-of-sight (NLOS) transmission due to obstacles \cite{5,105} severely degrades the performance of traditional subspace-based DOA estimation algorithms, including estimation of signal parameters via rotational invariance techniques (ESPRIT) \cite{6,7} and multiple signal classification (MUSIC) \cite{8,9}. With its low cost and flexible spatial reconfiguration capability, reconfigurable intelligent surface (RIS) \cite{10}, which is typically a planar array comprising subwavelength unit cells with controllable electromagnetic (EM) properties, can provide an alternative NLOS propagation path. By manipulating the phase, amplitude, and frequency of waves, RIS enables positioning and sensing in wireless systems \cite{11,12,Tuo3,13,14,Tuo1,106} in NLOS scenarios.
	
	There have been extensive studies on RIS-assisted localization of far-field (FF) targets. Regarding the performance of RIS-aided positioning, \cite{15} analyzes the impact of the number of RIS elements and phase configuration on the Cram\'er-Rao bound (CRB) for FF point localization in single-input multiple-output (MIMO) systems.  To fully exploit the adjustable phase advantage of RIS, a manifold optimization method was employed in \cite{16} to derive the optimal phase configuration by optimizing the CRB. For DOA estimation of FF targets, \cite{17} proposed a sparse RIS-based algorithm to address the NLOS problem in passive sensing systems, and \cite{18} developed a reduced-dimension MUSIC estimator for RIS-assisted MIMO radar systems. Beyond traditional subspace-based algorithms, sparse representation methods with RIS for DOA estimation have been explored in \cite{19,20,21,22,23}. In \cite{19}, a compressed sensing-based RIS channel estimator was proposed. In \cite{20}, a discrete-continuous optimization framework for orthogonal matching pursuit was formulated to mitigate off-grid errors. In \cite{21}, an individual channel estimation method based on atomic norm minimization (ANM) was established. To reduce complexity, a two-dimensional (2-D) angle estimation method using decoupled ANM (DANM) was proposed in \cite{22}, and DANM based on the alternating direction method of multipliers (ADMM) was further created in \cite{23}.
	
	All these methods assume that the target is located in the FF region. With the increase of antenna aperture and the decrease of signal wavelength, future 6G wireless systems are expected to operate in the radiated near-field (NF) region. The Rayleigh distance \cite{24}, which defines the boundary between FF and NF regions, will become increasingly large, and communication between users and RIS will frequently occur in the NF region. Consequently, the classic FF plane wave propagation model becomes invalid \cite{25,26}, rendering RIS-aided FF DOA estimation inapplicable for NF targets.
	
	To address the spatial propagation model of NF targets, several RIS-assisted NF localization algorithms have been proposed based on the maximum likelihood approach \cite{27,28,29}. {The work \cite{27} proposes a joint estimation algorithm to mitigate model mismatch for RIS-aided NF localization under phase-dependent amplitude variations.} The paper \cite{28} employed an iterative entropy-regularization-based phase design to focus the RIS phase configuration within a region of interest for multi-target NF localization. The article \cite{29} applied a two-stage localization approach together with 2-D signal-path classification to reduce the complexity of direct localization. \textbf{A critical limitation of these existing methods is that they all assume the RIS position to be known \textit{a priori}.} When the RIS location is unknown, the BS cannot directly establish the geometric relationship between itself and NF targets, making effective localization extremely challenging.
	
	Eliminating the requirement of known RIS position offers significant practical advantages for future 6G networks. First, it enables flexible and rapid RIS deployment without the need for precise positioning infrastructure or labor-intensive calibration procedures, which is particularly valuable for emergency communication scenarios, temporary network densification, or unmanned aerial vehicle (UAV)-mounted RIS systems where position information is inherently uncertain. Second, it substantially reduces the system deployment cost and maintenance complexity, as position monitoring hardware and periodic recalibration become unnecessary. Third, it enhances system robustness against environmental changes, as physical displacement of RIS due to wind, structural deformation, or deliberate relocation will not compromise localization capability. Most importantly, achieving accurate localization without the prior knowledge of RIS position fundamentally expands the application scope of RIS-aided sensing from controlled indoor environments to dynamic outdoor scenarios where RIS position information cannot be guaranteed.
	
	To overcome this fundamental limitation, we leverage the hybrid RIS (HRIS) \cite{30}, a recently emerged metamaterial surface architecture capable of simultaneously reflecting and receiving signals. From a signal processing perspective, the key advantage of HRIS lies in creating additional observability: HRIS receives signals from NF targets, which provides observations for target parameter estimation; and at the same time, simultaneously reflects signals to BSs, which yields observations for HRIS self-localization. This dual functionality fundamentally alters the identifiability conditions compared to pure reflection RIS, where the BS-to-target channel is a cascade of two unknown links (BS-RIS and RIS-target), making it impossible to separate target location from RIS location without prior knowledge of either one of them. With HRIS, the direct HRIS-to-target observations break this coupling, enabling a novel {three-step} paradigm: {NF target parameters are first estimated relative to the HRIS, the HRIS position is then determined through its received signals from BSs, and finally the NF target positions are converted to absolute coordinates via the now-known HRIS position.}
	
	However, realizing this paradigm poses several significant technical challenges. First, the NF signal model exhibits strong coupling between angle and range parameters due to spherical wavefront curvature, making it difficult to decouple and sequentially estimate these parameters. The commonly used Fresnel approximation simplifies the model but introduces systematic errors that degrade estimation accuracy, particularly for targets at close range. Second, HRIS self-localization requires estimating its position from the received BS signals, which involves solving a complex geometric triangulation problem with potential ambiguity issues when only angle information is available. Third, the phase optimization for HRIS must be performed without knowing the optimal geometric configuration, as the HRIS position itself is being estimated, creating a circular-dependency problem. Finally, achieving single-snapshot angle estimation for virtual FF models requires gridless sparse representation methods that can operate efficiently on high-dimensional covariance matrices without assuming multiple snapshots for spatial smoothing.
	
	In this paper, we propose a two-stage gridless NF localization framework assisted by position-unknown HRIS that systematically addresses these challenges. The first stage focuses on estimating NF target parameters relative to the HRIS and achieving HRIS self-localization, while the second stage leverages phase optimization to enhance localization accuracy. The main contributions are summarized as follows:
	\begin{itemize}
		\item \textbf{Parameter Estimation:} We propose a single-snapshot DANM to estimate the 2-D angles of the virtual FF model obtained from delay-based cross-correlation calculation of NF observation relative to HRIS, followed by a subspace method for range estimation.
		
		\item \textbf{Parameter Enhancement:} We develop a total least squares (TLS)-based parameter correction approach with the accurate NF model, which eliminates system errors introduced by the virtual FF model based on the Fresnel approximation.
		
		\item \textbf{Localization Algorithm:} We develop a least-squares (LS)  triangulation approach to determine the 3-D position of HRIS by utilizing the ANM method to solve the angle information of HRIS relative to two base stations.
		
		\item \textbf{Phase Optimization:} We propose a multi-stage phase design method, which uses semidefinite relaxation to discard the nonconvex constraint, and Gaussian randomization to obtain the HRIS phase matrix.
		
		\item \textbf{Performance Analysis:} We derive the CRB for the NF target parameters and the position error bound (PEB) for HRIS as performance benchmarks.
		
	\end{itemize}
	
	The remainder of this paper is organized as follows. Section II describes the HRIS-aided SIMO system for the NF signal. Section III introduces the proposed method, including virtual FF model construction, NF parameter estimation and correction, HRIS position estimation, as well as HRIS phase optimization. Section IV derives the theoretical CRB and PEB performance bounds. Section V presents comprehensive simulation results and analysis. Section VI draws a conclusion of the paper.
	\begin{figure*}[!t]
		\centering
		\centerline{\includegraphics[width=13 cm]{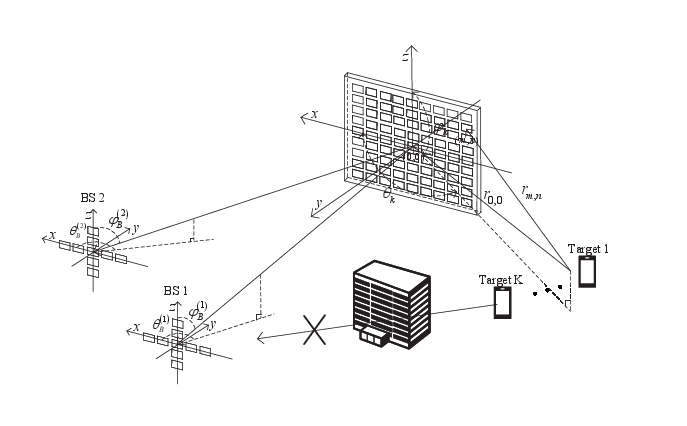}}
		\caption{HRIS-assisted NF target localization model.}
		\label{fig1}
	\end{figure*}
	Notations: Matrices and vectors are represented by bold uppercase and lowercase letters, respectively. ${\left(  \cdot  \right)^{\text{T}}}$, ${\left(  \cdot  \right)^*}$ and ${\left(  \cdot  \right)^{\text{H}}}$ correspond to transpose, conjugate, and conjugate transpose. $E\left\{  \cdot  \right\}$, $\left\|  \cdot  \right\|$ and $\left|  \cdot  \right|$ represent expectation, ${l_2}{\text{ - norm}}$, and absolute value. ${\left[ {\mathbf{a}} \right]_i}$ and ${\left[ {\mathbf{A}} \right]_{i,j}}$ are the $i$-th element of vector ${\mathbf{a}}$, and the $i$-th row and $j$-th column elements of matrix ${\mathbf{A}}$, respectively. ${\text{vec}}\left(  \cdot  \right)$ and ${\text{tr}}\left(  \cdot  \right)$ denote the matrix vectorization operation and the summation of matrix diagonal elements. $ \odot $, $ \otimes $ and $ \oplus $ are the Hadamard product, Kronecker product, and Khatri Rao product operations.
	
	\section{System Model}
	As shown in Fig. \ref{fig1}, we consider HRIS-assisted NF localization under the single-input multiple-output (SIMO) transmission model. The first BS at position ${\mathbf{p}}_B^{\left( 1 \right)} \in \mathbb{R}{}^3$ and the second BS at position ${\mathbf{p}}_B^{\left( 2 \right)} \in \mathbb{R}{}^3$ are both equipped with cross uniform linear arrays (ULA) with ${N_x}$ elements on the x-axis and ${N_z}$ elements on the z-axis. The HRIS at the unknown position ${{\mathbf{p}}_R} \in \mathbb{R}{}^3$ is positioned in the xoz plane, with a uniform planar array (UPA) structure having $M = \left( {2{m_x} + 1} \right) \times \left( {2{n_z} + 1} \right)$ elements, where ${m_x}$ is the number of elements along the positive x-axis, and ${n_z}$ is that along the positive z-axis. There are $K$ UE at unknown positions ${{\mathbf{p}}_U} \in \mathbb{R}{}^3$.  They are located in the NF region of the HRIS, and the direct path between a BS and each UE is blocked. Therefore, the two BSs receive signals from $K$ UEs through the reflected signals from the HRIS. At the $t$-th time slot, $t \in \left\{ {1,2, \cdots ,T} \right\}$, the received signal in the $\left( {m,n} \right)$-th HRIS element can be represented as
	
	\begin{equation}\label{eq1}
		\begin{aligned}
			y_{m,n}^{}\left( t \right) & = \sqrt \delta  \sum\limits_{k = 1}^K {{a_{m,n}}\left( {{\varphi _k},{\theta _k},{r_k}} \right)s_k^{}\left( t \right)}  + {\mathbf{\zeta }}_{m,n}^{}\left( t \right) , \\
			m & \in\{-m_x, \cdots, m_x \} , \;\; n\in\{-n_z, \cdots, n_z \} .
		\end{aligned}
	\end{equation}
	In \eqref{eq1}, ${\varphi _k}$, ${\theta _k}$, and ${r_k}$ are the elevation, azimuth, and range of the $k$-th target relative to the HRIS, respectively, and ${\varphi _k} = \arccos \left( {{{{{\left[ {{{\mathbf{p}}_U} - {{\mathbf{p}}_R}} \right]}_3}} \mathord{\left/
				{\vphantom {{{{\left[ {{{\mathbf{p}}_U} - {{\mathbf{p}}_R}} \right]}_3}} {\left\| {{{\mathbf{p}}_U} - {{\mathbf{p}}_R}} \right\|}}} \right.
				\kern-\nulldelimiterspace} {\left\| {{{\mathbf{p}}_U} - {{\mathbf{p}}_R}} \right\|}}} \right)$, ${\theta _k} = \arctan \left( {{{{{\left[ {{{\mathbf{p}}_U} - {{\mathbf{p}}_R}} \right]}_1}} \mathord{\left/
				{\vphantom {{{{\left[ {{{\mathbf{p}}_U} - {{\mathbf{p}}_R}} \right]}_1}} {{{\left[ {{{\mathbf{p}}_U} - {{\mathbf{p}}_R}} \right]}_2}}}} \right.
				\kern-\nulldelimiterspace} {{{\left[ {{{\mathbf{p}}_U} - {{\mathbf{p}}_R}} \right]}_2}}}} \right)$, ${r_k} = \left\| {{{\mathbf{p}}_U} - {{\mathbf{p}}_R}} \right\|$. $\delta$ is the common power allocation factor. ${\mathbf{\zeta }}_{m,n}^{}\left( t \right)$ is Gaussian white noise with zero mean and variance $\sigma _n^2$ at the $\left( {m,n} \right)$-th element. $s_k^{}\left( t \right)$ is the NF source signal. ${a_{m,n}}\left( {{\varphi _k},{\theta _k},{r_k}} \right) = {e^{j\frac{{2\pi }}{\lambda }\Delta r_{m,n}^k}}$ represents the ${m,n}$-th  element of the steering vector, where $\Delta r_{m,n}^k$ is the k-th signal path difference relative to the HRIS element located at $\left( {0,0} \right)$,
	\begin{equation}\label{eq2}
		\begin{aligned}
			\Delta r_{m,n}^k = r_{m,n}^k - r_{0,0}^k
		\end{aligned}
	\end{equation}
	$\lambda$ is the signal wavelength, and $r_{m,n}^k = \sqrt {{{\left( {r_{0,0}^k} \right)}^2} - 2dr_{0,0}^k{\omega _k} + \left( {{m^2} + {n^2}} \right){d^2}}$ is the distance between the $k$-th UE and the $\left( {m,n} \right)$-th element. ${\omega _k} = m\sin {\varphi _k}\cos {\theta _k} + n\cos {\varphi _k}$. Using the Fresnel approximation{\cite{107}}, (\ref{eq2}) becomes
	\begin{equation}\label{eq3}
		\begin{aligned}
			\Delta r_{m,n}^k \approx g_x^km + g_z^kn + \phi _x^k{m^2} + \phi _z^k{n^2} + {\alpha ^k}mn
		\end{aligned}
	\end{equation}
	where $g_x^k =  - d\sin {\varphi _k}\cos {\theta _k}$, $g_z^k =  - d\cos {\varphi _k}$, $\phi _x^k = \frac{{{d^2}}}{{2r_{0,0}^k}}\left( {1 - {{\sin }^2}{\varphi _k}{{\cos }^2}{\theta _k}} \right)$, $\phi _z^k = \frac{{{d^2}}}{{2r_{0,0}^k}}{\sin ^2}{\varphi _k}$, ${\alpha ^k} =  - \frac{{{d^2}}}{{2r_{0,0}^k}}\cos {\theta _k}\sin 2{\varphi _k}$, the element spacing $d = \frac{\lambda }{4}$. Hence, we have ${a_{m,n}}\left( {{\varphi _k},{\theta _k},{r_k}} \right) \approx {e^{j\frac{{2\pi }}{\lambda }\left( {g_x^km + g_z^kn + \phi _x^k{m^2} + \phi _z^k{n^2} + {\alpha ^k}mn} \right)}}$. The collection of all elements in \eqref{eq1} forms the HRIS received signal matrix
	\begin{equation}\label{eq4}
		\scalebox{0.9}{$\displaystyle
			\mathbf{Y}(t) =
			\begin{bmatrix}
				y_{-m_x,-n_z}(t) & \cdots & y_{-m_x, n_z-1}(t) & y_{-m_x, n_z}(t) \\
				y_{-m_x+1,-n_z}(t) & \cdots & \cdots & \vdots  \\
				\vdots & \ddots & \ddots & y_{m_x-1,n_z}(t) \\
				y_{m_x,-n_z}(t) & \cdots & y_{m_x, n_z-1}(t) & y_{m_x, n_z}(t)
			\end{bmatrix} .
			$}
	\end{equation}
	
	We now turn to the received signal at BS. The received signals of the two ULAs along the x-axis and z-axis at the $t$-th time slot for the $i$-th BS can be expressed as
	\begin{equation}\label{eq5}
		\begin{aligned}
			{\mathbf{y}}_{B,x}^{\left( i \right)}\left( t \right) = \sqrt {1 - \delta } {\mathbf{H}}_x^{\left( i \right)}{\mathbf{W}}_{x,t}^{\left( i \right)}{\mathbf{\bar y}}\left( t \right) + {\mathbf{\zeta }}\left( t \right), \quad i = 1, 2
		\end{aligned}
	\end{equation}
	\begin{equation}\label{eq6}
		\begin{aligned}
			{\mathbf{y}}_{B,z}^{\left( i \right)}\left( t \right) = \sqrt {1 - \delta } {\mathbf{H}}_z^{\left( i \right)}{\mathbf{W}}_{z,t}^{\left( i \right)}{\mathbf{\bar y}}\left( t \right) + {\mathbf{\zeta }}\left( t \right), \quad i = 1, 2
		\end{aligned}
	\end{equation}
	where ${\mathbf{\bar y}}\left( t \right) = {\text{vec}}\left( {{\mathbf{Y}}\left( t \right)} \right)$.  ${\mathbf{W}}_{x,t}^{\left( i \right)}$ represents the phase matrix of the HRIS for the received signal on the x-axis of the $i$-th BS at the $t$-th time slot.  It is given by ${\mathbf{W}}_{x,t}^{\left( i \right)} = {\text{diag}}\left( {{\mathbf{w}}_{x,t}^{\left( i \right)}} \right)$, where ${\mathbf{w}}_{x,t}^{\left( i \right)} = {\left[ {{e^{j{\psi _{ - {m_x}, - {n_z}}}}},{e^{j{\psi _{ - {m_x} + 1, - {n_z}}}}}, \cdots ,{e^{j{\psi _{{m_x},{n_z}}}}}} \right]^{\text{T}}} \in {\mathbb{C}^M}$. The other phase matrix ${\mathbf{W}}_{z,t}^{\left( i \right)}$ has the same form as ${\mathbf{W}}_{x,t}^{\left( i \right)}$.  ${\mathbf{H}}_x^{\left( i \right)}$ and ${\mathbf{H}}_z^{\left( i \right)}$ are the spatial response matrices of the HRIS on the x-axis and z-axis, respectively, of the $i$-th BS, which can be expressed as
	\begin{equation}\label{eq7}
		\begin{aligned}
			{\mathbf{H}}_x^{\left( i \right)} = {\gamma _x}{{\mathbf{b}}_x}\left( {\varphi _B^{\left( i \right)},\theta _B^{\left( i \right)}} \right)*{\mathbf{b}}_R^{\text{H}}\left( {\varphi _R^{\left( i \right)},\theta _R^{\left( i \right)}} \right) ,
		\end{aligned}
	\end{equation}
	\begin{equation}\label{eq8}
		\begin{aligned}
			{\mathbf{H}}_z^{\left( i \right)} = {\gamma _z}{{\mathbf{b}}_z}\left( {\varphi _B^{\left( i \right)}} \right)*{\mathbf{b}}_R^{\text{H}}\left( {\varphi _R^{\left( i \right)},\theta _R^{\left( i \right)}} \right) ,
		\end{aligned}
	\end{equation}
	\begin{figure}[!t]
		\centering
		\centerline{\includegraphics[width=7.5cm]{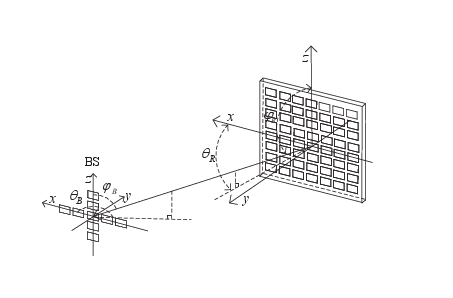}}
		\caption{Geometric relationship of BS-RIS path.}
		\label{fig2}
	\end{figure}
	and ${\gamma _x}$ and ${\gamma _z}$ are the path gains along the x-axis and z-axis from the HRIS to the BS. Here, $\varphi _B^{\left( i \right)}$ denotes the elevation and $\theta _B^{\left( i \right)}$ the azimuth angles of the HRIS relative to the $i$-th BS. The corresponding steering vectors are ${{\mathbf{b}}_x}\left( {\varphi _B^{\left( i \right)},\theta _B^{\left( i \right)}} \right) = {\left[ {{e^{j\frac{{\left( {N - 1} \right)}}{2}\sin \varphi _B^{\left( i \right)}\cos \theta _B^{\left( i \right)}}}, \cdots ,1, \cdots ,{e^{ - j\frac{{\left( {N - 1} \right)}}{2}\sin \varphi _B^{\left( i \right)}\cos \theta _B^{\left( i \right)}}}} \right]^{\text{T}}}$ and ${{\mathbf{b}}_z}\left( {\varphi _B^{\left( i \right)}} \right) = {\left[ {{e^{j\frac{{\left( {N - 1} \right)}}{2}\cos \varphi _B^{\left( i \right)}}}, \cdots ,1, \cdots ,{e^{ - j\frac{{\left( {N - 1} \right)}}{2}\cos \varphi _B^{\left( i \right)}}}} \right]^{\text{T}}}$. Furthermore,
	\begin{equation}\label{eq9}
		\begin{aligned}
			{{\mathbf{b}}_R}\left( {\varphi _R^{\left( i \right)},\theta _R^{\left( i \right)}} \right) = {{\mathbf{b}}_{R,z}}\left( {\varphi _R^{\left( i \right)}} \right) \otimes {{\mathbf{b}}_{R,x}}\left( {\theta _R^{\left( i \right)}} \right)
		\end{aligned}
	\end{equation}
	where \({{\mathbf{b}}_{R,x}}\left( {\theta _R^{\left( i \right)}} \right) = \left[ {e^{j{m_x}\sin \left( {\varphi _R^{\left( i \right)}} \right)\cos \left( {\theta _R^{\left( i \right)}} \right)}}, \cdots ,1, \cdots ,\right.\)
	\(\left.{e^{ - j{m_x}\sin \left( {\varphi _R^{\left( i \right)}} \right)\cos \left( {\theta _R^{\left( i \right)}} \right)}} \right]\), \({{\mathbf{b}}_{R,z}}\left( {\varphi _R^{\left( i \right)}} \right) = \left[ {e^{j{n_z}\cos \varphi _R^{\left( i \right)}}},\cdots\right.\)
	\(\left.,1,\cdots ,{e^{ - j{n_z}\cos \left( {\varphi _R^{\left( i \right)}} \right)}} \right]\). As illustrated in Fig. \ref{fig2}, $\varphi _R^{\left( i \right)}$ and $\theta _R^{\left( i \right)}$ are the elevation and azimuth angles from the $i$-th BS to the HRIS, respectively. Therefore, we have $\varphi _R^{\left( i \right)} = \pi  - \varphi _B^{\left( i \right)}$ and $\theta _R^{\left( i \right)} = \pi  - \theta _B^{\left( i \right)}$.
	
	Based on the above system model, the fundamental problem we aim to address is: \textbf{How to accurately locate the NLOS NF targets when the HRIS position ${{\mathbf{p}}_R}$ is unknown?} From a signal processing perspective, this problem presents several interrelated challenges that can be analyzed through the lens of parameter identifiability, Fisher information, and observability.
	
	\textbf{Challenge 1 (NF parameter coupling):} Estimating the 3-D NF target parameters $\left\{ {{\varphi _k},{\theta _k},{r_k}} \right\}$ from ${\mathbf{Y}}\left( t \right)$ in (\ref{eq1}) is complicated by the nonlinear coupling in the NF steering vector ${a_{m,n}}\left( {{\varphi _k},{\theta _k},{r_k}} \right) = {e^{j\frac{{2\pi }}{\lambda }\left( {g_x^km + g_z^kn + \phi _x^k{m^2} + \phi _z^k{n^2} + {\alpha ^k}mn} \right)}}$, where the linear terms (${g_x^k},{g_z^k}$) encode the angles, while the quadratic terms (${\phi _x^k},{\phi _z^k}$) encapsulate the range. The Fisher information for range is inherently lower than that for angles due to weaker phase sensitivity of the quadratic terms, leading to asymmetric estimation performance. Moreover, the Fresnel approximation in (\ref{eq3}) introduces a model mismatch that manifests as systematic bias, particularly affecting high-SNR performance.
	
	\textbf{Challenge 2 (HRIS observability):} Determining ${{\mathbf{p}}_R}$ from the BS observations in (\ref{eq5})-(\ref{eq6}) requires the extraction of HRIS-to-BS angles $\left\{ {\varphi _B^{\left( i \right)},\theta _B^{\left( i \right)}} \right\}$ from the cascaded channel ${\mathbf{H}}_x^{\left( i \right)} = {\gamma _x}{{\mathbf{b}}_x}\left( {\varphi _B^{\left( i \right)},\theta _B^{\left( i \right)}} \right)*{\mathbf{b}}_R^{\text{H}}\left( {\varphi _R^{\left( i \right)},\theta _R^{\left( i \right)}} \right)$. This is a low-rank matrix recovery problem, where angles are embedded in the Kronecker structure. With only two BSs, the geometric configuration must avoid collinearity to ensure triangulation uniqueness; otherwise, the position estimate suffers from rank deficiency.
	
	\textbf{Challenge 3 (Coordinate transformation):} Converting the HRIS-centric parameters to the BS-centric absolute coordinates is a nonlinear transformation that propagates estimation errors. The error propagation is governed by the Jacobian matrix ${\mathbf{J}} = \left[ {\frac{{\partial \theta _B^{\left( 1 \right)}}}{{\partial {{\mathbf{p}}_R}^{}}},\frac{{\partial \varphi _B^{\left( 1 \right)}}}{{\partial {{\mathbf{p}}_R}^{}}}} \right]$, whose conditioning directly impacts the position error bound (PEB).
	
	\textbf{Challenge 4 (Phase optimization without position):} Optimizing ${\mathbf{W}}_{x,t}^{\left( i \right)}$ to maximize received SNR requires knowledge of the channel ${\mathbf{H}}_x^{\left( i \right)}$, which depends on the unknown ${{\mathbf{p}}_R}$. This creates a circular dependency: better phase design improves HRIS localization (by increasing the Fisher information), which in turn enables better phase design. Breaking this loop requires iterative refinement or robust initialization.
	
	These challenges are fundamentally intertwined through the Fisher information matrix structure and necessitate a carefully designed two-stage framework that systematically addresses identifiability, bias correction, and information-theoretic optimization.
	
	\section{Proposed Algorithm}
	To tackle the aforementioned challenges under the position-unknown HRIS scenario, we propose a systematic two-stage gridless localization framework that embodies the {aforementioned three-step} paradigm. In the first stage, we estimate the NF target 3-D parameters $\left\{ {{\varphi _k},{\theta _k},{r_k}} \right\}$ relative to the HRIS, and {determine the HRIS position ${{\mathbf{p}}_R}$ through its received signals from two BSs. In the second stage, we optimize the HRIS phase configuration to enhance localization accuracy and convert the target positions to the absolute coordinates. The detailed algorithmic procedures are presented in the following subsections.
		\subsection{Single snapshot virtual FF data construction}
		Due to the high coupling between the angle and range of NF targets in the steering vector of the HRIS, efficiently estimating the angles requires eliminating the influence of quadratic and cross terms related to ${r_k}$ in ${a_{m,n}}\left( {{\theta _k},{\varphi _k},{r_k}} \right)$. At time delay $\tau$, performing cross-correlation on ${\mathbf{\bar y}}\left( t \right)$ gives
		\begin{equation}\label{eq10}
			\begin{aligned}
				{h_{{l_{m,n}},{l_{ - m, - n}}}}\left( \tau  \right) &= E\left\{ {{{\left[ {{\mathbf{\bar y}}\left( {t + \tau } \right)} \right]}_u}{{\left[ {{\mathbf{\bar y}}_{}^*\left( t \right)} \right]}_v}} \right\} \\
				&= {\mathbf{a}}_x^{\text{T}}\left( m \right){\text{diag(}}{\mathbf{s}}\left( \tau  \right){\text{)}}{\mathbf{a}}_z^{}\left( n \right) , \\
			\end{aligned}
		\end{equation}
		where ${{l}_{\ast,\circ}}$ represents the $\left( {\ast,\circ} \right)$-th element of the HRIS, $u = \left( {2{m_x} + 1} \right)\left( {n + {n_z}} \right) + \left( {m + {m_x} + 1} \right)$, $a_m^{}({\varphi _k},{\theta _k}) = {e^{j\frac{{4\pi }}{\lambda }g_x^km}}$, $a_n^{}({\varphi _k}) = {e^{j\frac{{4\pi }}{\lambda }g_z^km}}$, ${{\mathbf{a}}_x}\left( m \right) = {\left[ {a_m^{}({\varphi _1},{\theta _1}),a_m^{}({\varphi _2},{\theta _2}), \cdots ,a_m^{}({\varphi _K},{\theta _K})} \right]^{\text{T}}}$, ${{\mathbf{a}}_z}\left( n \right) = {\left[ {a_n^{}({\varphi _1}),a_n^{}({\varphi _2}), \cdots ,a_n^{}({\varphi _K})} \right]^{\text{T}}}$, {and} {\({\mathbf{s}}\left( \tau  \right) = \delta  \left[ E\left\{ {{s_1}\left( {t + \tau } \right)s_1^*\left( t \right)} \right\},E\left\{ {{s_2}\left( {t + \tau } \right)s_2^*\left( t \right)} \right\}, \cdots ,E\left\{ {s_K}\left( t \right.\right.\right.\)
			\(\left.\left.\left.+ \tau  \right)s_K^*\left( t \right) \right\} \right]^{\text{T}}\)}.
		
		By traversing all elements of the HRIS array, ${h_{{{l}_{m,n}},{{l}_{ - m, - n}}}}\left( \tau  \right)$ can be written in matrix form as
		\begin{equation}\label{eq11}
			\scalebox{0.90}{%
				$\displaystyle
				\begin{aligned}
					\mathbf{H}(\tau)
					&= \begin{bmatrix}
						h_{l_{-m_x,-n_z},l_{m_x,n_z}}(\tau) & \cdots & h_{l_{-m_x,n_z},l_{m_x,-n_z}}(\tau) \\
						\vdots & \ddots & \vdots \\
						h_{l_{m_x,-n_z},l_{-m_x,n_z}}(\tau) & \cdots & h_{l_{m_x,n_z},l_{-m_x,-n_z}}(\tau)
					\end{bmatrix} \\
					&= \mathbf{A}_{\text{x}}^{\text{T}} \operatorname{diag}\left( \mathbf{s}(\tau) \right) \mathbf{A}_z
				\end{aligned}$%
			}
		\end{equation}
		where ${{\mathbf{A}}_x} = {\left[ {{{\mathbf{a}}_x}\left( { - {m_x}} \right),{{\mathbf{a}}_x}\left( { - {m_x} + 1} \right), \cdots ,{{\mathbf{a}}_x}\left( {{m_x}} \right)} \right]^{\text{T}}}$, ${{\mathbf{A}}_z} = {\left[ {{{\mathbf{a}}_z}\left( { - {n_z}} \right),{{\mathbf{a}}_z}\left( { - {n_z} + 1} \right), \cdots ,{{\mathbf{a}}_z}\left( {{n_z}} \right)} \right]^{\text{T}}}$.
		
		Vectorizing ${\mathbf{H}}\left( \tau  \right)$ yields
		\begin{equation}\label{eq12}
			\begin{aligned}
				{\mathbf{h}}\left( \tau  \right) &= {\text{vec}}\left( {{\mathbf{H}}\left( \tau  \right)} \right) \\
				&= \sum\limits_{k{\text{ = 1}}}^K {{\mathbf{\tilde a}}_{{n_z}}^{}\left( {{\varphi _k}} \right) \otimes {\mathbf{\tilde a}}_{{m_x}}^{}\left( {{\varphi _k},{\theta _k}} \right){{\left[ {{\mathbf{s}}\left( \tau  \right)} \right]}_k}}  \\
			\end{aligned}
		\end{equation}
		where \({{\mathbf{\tilde a}}_{{m_x}}}\left( {{\varphi _k},{\theta _k}} \right) = \left[ a_{ - {m_x}}^{}({\varphi _k},{\theta _k}),a_{ - {m_x} + 1}^{}({\varphi _k},{\theta _k}), \cdots ,\right.\)
		\(\left.a_{{m_x}}^{}({\varphi _k},{\theta _k}) \right]^{\text{T}}\), \({\mathbf{\tilde a}}_{{n_z}}^{}\left( {{\varphi _k}} \right) = \left[ a_{ - {n_z}}^{}({\varphi _k}),a_{ - {n_z} + 1}^{}({\varphi _k}), \cdots \right.\)
		\(\left. ,a_{{n_z}}^{}({\varphi _k}) \right]^{\text{T}}\).
		
		For lags $\tau  \in \left\{ {{\tau _0},{\text{ }}{\tau _1}, \cdots ,{\text{ }}{\tau _{L-1}}} \right\}$, vectors ${\mathbf{h}}\left( \tau  \right)$ are collected in a matrix of size $M \times L$,
		\begin{equation}\label{eq13}
			\begin{aligned}
				{\mathbf{\tilde H}} = \left[ {{\mathbf{h}}\left( {{\tau _0}} \right),{\mathbf{h}}\left( {{\tau _1}} \right), \cdots ,{\mathbf{h}}\left( {{\tau _{L-1}}} \right)} \right]
			\end{aligned}
		\end{equation}
		where $L$ is the number of pseudo snapshots.
		
		To expand the virtual array aperture of the HRIS, the covariance computation and vectorization of ${\mathbf{\tilde H}}$ can be obtained as
		\begin{equation}\label{eq14}
			\begin{aligned}
				{\mathbf{\tilde z}} &= {\text{vec}}\left( {{\text{E}}\left\{ {{\mathbf{\tilde H}}*{{{\mathbf{\tilde H}}}^{\text{H}}}} \right\}} \right) \\
				&= \left( {{{{\mathbf{\tilde A}}}^*} \oplus {\mathbf{\tilde A}}} \right){\mathbf{q}} \\
			\end{aligned}
		\end{equation}
		where ${\mathbf{\tilde A}} = \left[ {{{{\mathbf{\bar a}}}_1},{{{\mathbf{\bar a}}}_2}, \cdots ,{{{\mathbf{\bar a}}}_K}} \right]$, ${{\mathbf{\bar a}}_k} = {\mathbf{\tilde a}}_{{n_z}}^{}\left( {{\varphi _k}} \right) \otimes {\mathbf{\tilde a}}_{{m_x}}^{}\left( {{\varphi _k},{\theta _k}} \right)$, ${q_k} = E\left\{ {{{\left[ {{\mathbf{s}}\left( \tau  \right)} \right]}_k}{{\left( {{{\left[ {{\mathbf{s}}\left( \tau  \right)} \right]}_k}} \right)}^{\text{H}}}} \right\}$, ${\mathbf{q}} = {\left[ {q_1^{},q_2^{}, \cdots ,q_K^{}} \right]^{\text{T}}}$. The proposed single-snapshot virtual FF data construction is summarized in Algorithm 1.
		
		\begin{table}[htbp]
			\centering
			\label{table1}
			\begin{tabular}{l p{0.43\textwidth}}
				\toprule
				\multicolumn{2}{l}{\textbf{Algorithm 1:} Single snapshot virtual FF data construction} \\
				\midrule
				\multicolumn{2}{l}{\textbf{input:} ${\mathbf{\bar y}}\left( t \right)$} \\
				\multicolumn{2}{l}{\textbf{output:} ${\mathbf{\tilde z}}$} \\
				1: & \textbf{for} $m \leftarrow  - {m_x}$ \textbf{to} $m_x$ \textbf{do}\\
				2: & \quad \textbf{for} $n \leftarrow  - {n_z}$ \textbf{to} $n_z$ \textbf{do} \\
				3: & \quad \quad Select the centrally symmetric array data from ${\mathbf{\bar y}}\left( t \right)$ for cross-\\
				& \quad \quad correlation calculation shown in (\ref{eq10}). \\
				4: & \quad \textbf{end} \\
				5: & \textbf{end} \\
				6: & Traverse all the elements of the HRIS array according to (\ref{eq11}) to obtain ${\mathbf{H}}\left( \tau  \right)$. \\
				7: & Vectorize $\mathbf{H}(\tau)$ to obtain ${\mathbf{h}}\left( \tau  \right)$ and collect vectors $\mathbf{h}(\tau)$ in a matrix ${\mathbf{\tilde H}}$ of size $M \times L$. \\
				8: & Vectorize ${\mathbf{\tilde H}}$ to obtain ${\mathbf{\tilde z}}$. \\
				\bottomrule
			\end{tabular}
		\end{table}
		\subsection{Parameter Estimation and Correction}
		In (\ref{eq14}), the virtual received data ${\mathbf{\tilde z}}$ is single-snapshot data. To estimate the 2-D angles of the NF source, we propose a DANM method for the single snapshot case, and then apply the TLS-based parameter correction approach with the accurate NF model (\ref{eq2}) to improve the estimation accuracy.

		Let us denote the set of atoms of the atomic norm as $\mathcal{A} = \left\{ {{\mathbf{\tilde a}}_{{{\bar n}_z}}^{}\left( \varphi  \right) \otimes {\mathbf{\tilde a}}_{{{\bar m}_x}}^{}\left( {\varphi ,\theta } \right):\theta  \in \left( {0,\pi } \right),\varphi  \in \left( {0,\pi } \right)} \right\}$, where ${\bar m_x}$ and ${\bar n_z}$ are, respectively, the virtual element dimensions along the x- and z-axes. Then, the atomic norm of ${\mathbf{z}}$ on the atomic set $\mathcal{A}$ is defined as \({\left\| {\mathbf{z}} \right\|_\mathcal{A}} = \mathop {\inf }\limits_{{\theta _k},{\varphi _k},q_k^2} \left\{ \sum\nolimits_{k = 1}^K {\left| {q_k^{}} \right|} :{\mathbf{z}} = \sum\nolimits_{k = 1}^K {\mathbf{\tilde a}}_{\bar n}^{}\left( {{\varphi _k}} \right) \otimes {\mathbf{\tilde a}}_{\bar m}^{}\left( {{\varphi _k},{\theta _k}} \right)q_k^{},\right.\)
		\(\left.{\text{ }}{\theta _k} \in \left( {0,\pi } \right),{\varphi _k} \in \left( {0,\pi } \right)  \right\}\)
		, which can be calculated via semidefinite programming (SDP) by
		\begin{equation}\label{eq15}
			\begin{gathered}
				\mathop {\min }\limits_{x{\text{,}}{u_{_1}}} \frac{{{\rho _R}}}{2}\left( {x + {u_1}} \right) + \left\| {{\mathbf{\tilde z}} - {\mathbf{z}}} \right\| \hfill \\
				{\text{s}}{\text{.t}}{\text{.}}\left[ {\begin{array}{*{20}{c}}
						x&{{{\mathbf{z}}^{\mathbf{H}}}} \\
						{\mathbf{z}}&{{{\mathbf{T}}_{x,z}}\left( {\mathbf{u}} \right)}
				\end{array}} \right] \geqslant 0 \hfill \\
			\end{gathered}
		\end{equation}
		where ${{\text{T}}_{x,z}}\left( {\text{u}} \right) \in {\mathbb{C}^{\bar m\bar n*\bar m\bar n}}$ is a two-level Toeplitz matrix, ${u_1}$ is the first column and the first row element of ${{\text{T}}_{x,z}}\left( {\text{u}} \right)$, ${\rho _R} = {\sigma _n}\sqrt {\bar m\bar n\log \left( {\bar m\bar n} \right)} $, and $x$ is a constant.
		
		Due to the rapidly growing complexity of the optimization problem in (\ref{eq15}) with the size of $\bar m$ and $\bar n$, DANM is introduced to separate the dimensions of the coupled 2-D angle information in HRIS, while all information in both dimensions is jointly utilized to maintain optimal performance.
		
		The decoupled atomic set is defined as ${\mathcal{A}_d} = \left\{ {{\mathbf{\tilde a}}_{{{\bar m}_x}}^{}\left( {\varphi ,\theta } \right){\mathbf{\tilde a}}_{{{\bar n}_z}}^{}{{\left( {\pi  - \varphi } \right)}^{\text{H}}}:\theta  \in \left( {0,\pi } \right),\varphi  \in \left( {0,\pi } \right)} \right\}$, and the atomic norm of ${\mathbf{z}}$ on the atomic set ${\mathcal{A}_d}$ is defined as \({\left\| {\mathbf{z}} \right\|_{{\mathcal{A}_d}}} = \mathop {\inf }\limits_{{\theta _k},{\varphi _k},q_k^2} \left\{ \sum\nolimits_{k = 1}^K {\left| {q_k^{}} \right|} :{{\mathbf{z}}_d} = \sum\nolimits_{k = 1}^K {\mathbf{\tilde a}}_{\bar m}^{}\left( {{\varphi _k},{\theta _k}} \right){\mathbf{\tilde a}}_{\bar n}^{}{{\left( {\pi  - {\varphi _k}} \right)}^{\text{H}}}q_k^{},\right.\)
		\(\left.{\text{ }}{\theta _k} \in \left( {0,\pi } \right),{\varphi _k} \in \left( {0,\pi } \right) \right\}\).  It follows from the notion of \eqref{eq15} that we have the following optimization problem:
		\begin{equation}\label{eq16}
			\begin{gathered}
				\mathop {\min }\limits_{\mathbf{z}} \rho \left( {\frac{1}{{2\bar n}}{\text{tr}}\left( {{{\mathbf{T}}_z}\left( {\mathbf{u}} \right)} \right) + \frac{1}{{2\bar m}}{\text{tr}}\left( {{{\mathbf{T}}_x}\left( {\mathbf{u}} \right)} \right)} \right) + \left\| {{\mathbf{\tilde z}} - {\text{vec}}\left( {{{\mathbf{z}}_d}} \right)} \right\| \hfill \\
				{\text{s}}{\text{.t}}{\text{.}}\left[ {\begin{array}{*{20}{c}}
						{{{\mathbf{T}}_z}\left( {\mathbf{u}} \right)}&{{\mathbf{z}}_d^{\text{H}}} \\
						{{{\mathbf{z}}_d}}&{{{\mathbf{T}}_x}\left( {\mathbf{u}} \right)}
				\end{array}} \right] \geqslant 0 \hfill \\
			\end{gathered}
		\end{equation}
		
		With (\ref{eq16}), $\hat \varphi $ and $\hat \phi $ can be obtained by the MUSIC algorithm from ${{\mathbf{T}}_x}\left( {\mathbf{u}} \right)$ and ${{\mathbf{T}}_z}\left( {\mathbf{u}} \right)$, respectively, where $\hat \phi  = \sin \left( {\hat \varphi } \right)\cos (\hat \theta )$. Since $\hat \varphi $ and $\hat \phi $ are not inherently paired, we propose an angle pairing algorithm to identify the $\left( {{{\hat \varphi }_i},{{\hat \phi }_j}} \right),\forall i,j \in \left\{ {1,2, \cdots ,K} \right\}$ by:
		\begin{equation}\label{eq17}
			\begin{aligned}
				j = \mathop {\arg \max }\limits_j {\left[ {{\mathbf{A}}_z^\dag {{\mathbf{z}}_d}{{\left( {{\mathbf{A}}_x^{\text{H}}} \right)}^\dag }} \right]_{i,j}},i = 1, \cdots ,K
			\end{aligned}
		\end{equation}
		
		After the correct $\left( {{{\hat \varphi }_i},{{\hat \phi }_j}} \right)$ is found, then we have ${\hat \theta _j} = \arcsin \left( {\frac{{{{\hat \phi }_j}}}{{\sin \left( {{{\hat \varphi }_i}} \right)}}} \right)$. The MUSIC algorithm is then used to obtain the range estimate $\hat r$ using the covariance matrix of ${\mathbf{\bar y}}\left( t \right)$ by substituting $\left( {{{\hat \varphi }_i},{{\hat \theta }_j}} \right)$ into ${a_{m,n}}\left( {{\varphi _k},{\theta _k},{r_k}} \right)$.
		
		To reduce algorithmic complexity while preserving the Vandermonde structure of the steering vector, the Fresnel approximation has been adopted for the original exact model.
		
		We next utilized the TLS method to mitigate systematic errors, through minimizing the parameter estimation deviation. Putting $\big( {\hat \varphi _k^{},\hat \theta _k^{},\hat r_k^{}} \big)$ obtained under the Fresnel approximation to (\ref{eq3}) yields $\Delta \hat r_{m,n}^k$. Then, substituting $\Delta \hat r_{m,n}^k$ into $\Delta r_{m,n}^k = r_{m,n}^k - r_{0,0}^k$ yields
		\begin{equation}\label{eq18}
			\begin{aligned}
				\scalebox{0.96}{$
					\displaystyle
					2dr_{0,0}^k{\omega _k} + 2\Delta \hat r_{m,n}^kr_{0,0}^k = \left( {{m^2} + {n^2}} \right){d^2} - {\left( {\Delta \hat r_{m,n}^k} \right)^2}$
				}
			\end{aligned}
		\end{equation}
		
		Considering $r_{0,0}^k\cos {\theta _k}\sin {\varphi _k}$, $r_{0,0}^k\cos {\varphi _k}$, and $r_{0,0}^k$ in (\ref{eq18}) as unknown variables, an overdetermined system of equations can be used to construct the matrix ${\mathbf{P}}$ as
		\begin{equation}\label{eq19}
			\scalebox{0.57}{$\displaystyle
				\begin{aligned}
					{\mathbf{P}} = \left[ {\begin{array}{*{20}{c}}
							{ - 2d{m_x}}&{ - 2d{n_z}}&{2\Delta \hat r_{ - {m_{\text{x}}}, - {n_z}}^k}&{ - \left( {m_x^2 + n_z^2} \right){d^2} + {{\left( {\Delta \hat r_{ - {m_{\text{x}}}, - {n_z}}^k} \right)}^2}} \\
							{2d\left( { - {m_x} + 1} \right)}&{ - 2d{n_z}}&{2\Delta \hat r_{ - {m_x} + 1, - {n_z}}^k}&{ - \left( {{{\left( { - m_x^{} + 1} \right)}^2} + n_z^2} \right){d^2} + {{\left( {\Delta \hat r_{ - {m_x} + 1, - {n_z}}^k} \right)}^2}} \\
							\vdots & \vdots & \vdots & \vdots  \\
							{2d{m_x}}&{2d{n_z}}&{2\Delta \hat r_{{m_x},{n_z}}^k}&{ - \left( {m_x^2 + n_z^2} \right){d^2} + {{\left( {\Delta \hat r_{{m_x},{n_z}}^k} \right)}^2}}
					\end{array}} \right]
				\end{aligned}
				$}
		\end{equation}
		
		Let ${\mathbf{f}} \in {\mathbb{R}^4}$ be the right singular vector associated with the minimum singular value of matrix ${\mathbf{P}}$, then we can obtain
		\begin{equation}\label{eq20}
			\begin{aligned}
				\hat \varphi _k^{\mathbf{f}} = {\text{arccos}}\left( {{{\left[ {\mathbf{f}} \right]}_2}/{{\left[ {\mathbf{f}} \right]}_3}} \right)
			\end{aligned}
		\end{equation}
		\begin{equation}\label{eq21}
			\begin{aligned}
				\hat \theta _k^{\mathbf{f}} = \arccos \left( {{{\left[ {\mathbf{f}} \right]}_1}/\left( {{{\left[ {\mathbf{f}} \right]}_3}\sin \hat \varphi _k^{\mathbf{f}}} \right)} \right)
			\end{aligned}
		\end{equation}
		\begin{equation}\label{eq22}
			\begin{aligned}
				\hat r_k^{\mathbf{f}} = {\left[ {\mathbf{f}} \right]_3}/{\left[ {\mathbf{f}} \right]_4}
			\end{aligned}
		\end{equation}
		
		The parameter estimation and correction methods for the NF targets are summarized in Algorithm 2.
		\begin{table}[htbp]
			\centering
			\label{table2}
			\begin{tabular}{l p{0.43\textwidth}}
				\toprule
				\multicolumn{2}{l}{\textbf{Algorithm 2:} Parameter Estimation and Correction for NF Targets} \\
				\midrule
				\multicolumn{2}{l}{\textbf{input:} ${\mathbf{\tilde z}}$} \\
				\multicolumn{2}{l}{\textbf{output:} $\hat \varphi _k^{\mathbf{f}}$, $\hat \theta _k^{\mathbf{f}}$, $\hat r_k^{\mathbf{f}}$} \\
				1: & Construct DANM based on ${\mathbf{\tilde z}}$, and obtain ${{\mathbf{T}}_x}\left( {\mathbf{u}} \right)$ and ${{\mathbf{T}}_z}\left( {\mathbf{u}} \right)$ using (\ref{eq16}). \\
				2: & Applying the MUSIC algorithm to ${{\mathbf{T}}_x}\left( {\mathbf{u}} \right)$ and ${{\mathbf{T}}_z}\left( {\mathbf{u}} \right)$ yields unpaired $\hat \phi $ and $\hat \varphi $. \\
				3: & \textbf{for} $i \leftarrow 1$ \textbf{to} $K$ \textbf{do}\\
				4: & \quad \textbf{for} $j \leftarrow 1$ \textbf{to} $K$ \textbf{do}\\
				5: & \quad \quad Compute $\mathop {\arg \max }\limits_j {\left[ {{\mathbf{A}}_z^\dag {{\mathbf{z}}_d}{{\left( {{\mathbf{A}}_x^{\text{H}}} \right)}^\dag }} \right]_{i,j}}$ to find the paired $\left( {{{\hat \varphi }_i},{{\hat \phi }_j}} \right)$.  \\
				6: & \quad \quad Calculate $\arcsin \left( {\frac{{{{\hat \phi }_j}}}{{\sin \left( {{{\hat \varphi }_i}} \right)}}} \right)$ to obtain ${\hat \theta _j}$.\\
				7: & \quad \textbf{end} \\
				8: & \textbf{end} \\
				9: & Calculate the covariance matrix ${\mathbf{\bar y}}\left( t \right)$ for the MUSIC algorithm to obtain $\hat r$. \\
				10: & \textbf{For} $k \leftarrow 1$ \textbf{to} $K$ \textbf{do}\\
				11: & \quad Obtain the approximate delay $\Delta \hat r_{m,n}^k$ from (\ref{eq3}).\\
				12:	& \quad Apply $\Delta \hat r_{m,n}^k$ to (\ref{eq2}), and then form the overdetermined \\
				& \quad system matrix ${\mathbf{P}}$ with (\ref{eq19}).\\
				13: & \quad Perform singular value decomposition (SVD) on ${\mathbf{P}}$ to obtain the \\
				& \quad right singular vector ${\mathbf{f}}$.\\
				14: &\quad $\hat \varphi _k^{\mathbf{f}} = {\text{arccos}}\left( {{{\left[ {\mathbf{f}} \right]}_2}/{{\left[ {\mathbf{f}} \right]}_3}} \right)$, $\hat \theta _k^{\mathbf{f}} = \arccos \left( {{{\left[ {\mathbf{f}} \right]}_1}/\left( {{{\left[ {\mathbf{f}} \right]}_3}\sin \hat \varphi _k^{\mathbf{f}}} \right)} \right)$,\\
				&\quad $\hat r_k^{\mathbf{f}} ={\left[ {\mathbf{f}} \right]_3}/{\left[ {\mathbf{f}} \right]_4}$.\\
				15: & \textbf{end} \\
				\bottomrule
			\end{tabular}
		\end{table}
		
		\begin{remark}[Complexity and Bias-Variance Tradeoff]
			The proposed DANM method in Algorithm 2 achieves computational efficiency through dimension decoupling while maintaining super-resolution capability. Specifically, the complexity of the coupled 2-D ANM (\ref{eq15}) scales as $\mathcal{O}\left( {{\bar m}^3{\bar n}^3} \right)$ due to the Toeplitz structure over the product space, whereas DANM reduces it to $\mathcal{O}\left( {{\bar m}^3 + {\bar n}^3} \right)$ by separately optimizing ${\mathbf{T}}_x$ and ${\mathbf{T}}_z$. This decoupling is justified by the observation that the virtual FF model (\ref{eq13}) exhibits separable structures along the x- and z-axes, in which DANM exploits without sacrificing estimation accuracy. {We note that this separable structure is a specific property arising from the cross-array configuration adopted in this work, where the x-axis and z-axis elements form independent subarrays. For general 2-D planar arrays (e.g., full UPA), the exact separability does not hold, and the DANM decoupling would introduce approximation loss in estimation accuracy.}
			
			Moreover, the TLS correction (Steps 10--15) addresses a fundamental bias-variance tradeoff. The Fresnel approximation introduces systematic bias that dominates the estimation error at high SNR, leading to an error floor. TLS mitigates this bias by refining the estimates under the exact NF model (\ref{eq2}). However, TLS relies on SVD of the overdetermined matrix ${\mathbf{P}}$, which amplifies noise when the system is ill-conditioned (e.g., when the targets are at large distances where $\Delta \hat r_{m,n}^k / \hat r_{0,0}^k \approx 0$). Consequently, TLS is most beneficial in moderate-to-high SNR regimes where bias dominates variance, as confirmed in Fig. \ref{fig5} where the performance gap between DANM and DANM+TLS widens with increasing SNR.
		\end{remark}
		
		\subsection{HRIS location estimation}
		To estimate the position of HRIS, we apply an LS-based geometric triangulation method by utilizing the positions of two BSs. To this end, we reconstruct ${\mathbf{\hat Y}}\left( t \right)$ with the estimate of $\hat \varphi _k^{\mathbf{f}}$, $\hat \theta _k^{\mathbf{f}}$, and $\hat r_k^{\mathbf{f}}$ according to (\ref{eq4}), and vectorize ${\mathbf{\hat Y}}\left( t \right)$ to have ${\mathbf{\hat y}}\left( t \right) = {\text{vec}}\left( {{\mathbf{\hat Y}}\left( t \right)} \right)$, then construct a new matrix ${{\mathbf{\hat Y}}_w}$ as follows,
		\begin{equation}\label{eq23}
			\begin{aligned}
				{{\mathbf{\hat Y}}_w} = \left[ {{\mathbf{w}}_{x,t}^{\left( 1 \right)} \odot {\mathbf{\hat y}}\left( 1 \right), \cdots ,{\mathbf{w}}_{x,T}^{\left( 1 \right)} \odot {\mathbf{\hat y}}\left( {\text{T}} \right)} \right]
			\end{aligned}
		\end{equation}
		
		Let us consider the received data from the first BS along the x-axis and let the atomic set $\mathcal{B}$ be $\mathcal{B} = \left\{ {{{\mathbf{b}}_x}\left( {\varphi ,\theta } \right){{\mathbf{b}}_R}{{\left( {\pi  - \varphi ,\pi  - \theta } \right)}^{\text{H}}}:\varphi  \in \left( {0,\pi } \right),\theta  \in \left( {0,\pi } \right)} \right\}$ according to (\ref{eq7}). The problem of ANM of ${{\mathbf{H}}_x}$ over the atomic set $\mathcal{B}$ is
		\begin{equation}\label{eq24}
			\begin{aligned}
				\min {\text{ }}{\left\| {{\mathbf{y}}_{B,x}^{\left( 1 \right)} - \sqrt {1 - \delta } {\mathbf{H}}_x^{\left( 1 \right)}{{{\mathbf{\hat Y}}}_w}} \right\|_F} + {\left\| {{\mathbf{H}}_x^{\left( 1 \right)}} \right\|_{\mathcal{B}}}
			\end{aligned}
		\end{equation}
		
		To solve for $\theta _B^{\left( 1 \right)}$, an equivalent semidefinite optimization problem can be constructed with (\ref{eq24}) as
		\begin{equation}\label{eq25}
			\scalebox{0.80}{$\displaystyle
				\begin{gathered}
					\mathop {\min }\limits_{{{\mathbf{H}}_x}} {\text{ }}{\left\| {{\mathbf{y}}_{B,x}^{\left( 1 \right)} - \sqrt {1 - \delta } {\mathbf{H}}_x^{\left( 1 \right)}{{{\mathbf{\hat Y}}}_w}} \right\|_F} + {\rho _B}\left( {\frac{1}{{2N}}{\text{tr}}\left( {{{\mathbf{T}}_{B,x}}\left( {\mathbf{u}} \right)} \right) + \frac{1}{{2M}}{\text{tr}}\left( {{{\mathbf{T}}_R}\left( {\mathbf{u}} \right)} \right)} \right) \hfill \\
					{\text{s}}{\text{.t}}{\text{.}}\left[ {\begin{array}{*{20}{c}}
							{{\mathbf{T}}_R^{\left( 1 \right)}\left( {\mathbf{u}} \right)}&{\left( {{\mathbf{H}}_x^{\left( 1 \right)}} \right)_{}^{\text{H}}} \\
							{{\mathbf{H}}_x^{\left( 1 \right)}}&{{\mathbf{T}}_{B,x}^{\left( 1 \right)}\left( {\mathbf{u}} \right)}
					\end{array}} \right] \geqslant 0 \hfill \\
				\end{gathered}
				$}
		\end{equation}
		where ${\rho _B} = {\sigma _n}\sqrt {NM\log \left( {NM} \right)} $, ${\mathbf{T}}_{B,x}^{\left( 1 \right)}\left( {\mathbf{u}} \right) = {{\mathbf{b}}_x}\left( {\varphi _B^{\left( 1 \right)},\theta _B^{\left( 1 \right)}} \right){{\mathbf{D}}_x}{\mathbf{b}}_x^{\text{H}}\left( {\varphi _B^{\left( 1 \right)},\theta _B^{\left( 1 \right)}} \right)$, and ${{\mathbf{D}}_x}$ is a positive semidefinite diagonal matrix.
		
		For the received data on the z-axis of the first BS, the corresponding ${\mathbf{T}}_{B,z}^{\left( 1 \right)}\left( {\mathbf{u}} \right)$ can also be calculated using a similar procedure as with (\ref{eq24})-(\ref{eq25}), where ${\mathbf{T}}_{B,z}^{\left( 1 \right)}\left( {\mathbf{u}} \right) = {{\mathbf{b}}_z}\left( {\varphi _B^{\left( 1 \right)}} \right){{\mathbf{D}}_z}{\mathbf{b}}_z^{\text{H}}\left( {\varphi _B^{\left( 1 \right)}} \right)$, and ${{\mathbf{D}}_z}$ is a positive semidefinite diagonal matrix. Then, we apply the MUSIC algorithm to ${\mathbf{T}}_{B,x}^{\left( 1 \right)}\left( {\mathbf{u}} \right)$ to solve for $\phi _B^{\left( 1 \right)}$ and ${\mathbf{T}}_{B,z}^{\left( 1 \right)}\left( {\mathbf{u}} \right)$ for $\varphi _B^{\left( 1 \right)}$, where $\phi _B^{\left( 1 \right)} = \sin \varphi _B^{\left( 1 \right)}\cos \theta _B^{\left( 1 \right)}$, and $\theta _B^{\left( 1 \right)} = \arccos \left( {\frac{{\phi _B^{\left( 1 \right)}}}{{\sin \left( {\varphi _B^{\left( 1 \right)}} \right)}}} \right)$. We can solve $\varphi _B^{\left( 2 \right)}$ and $\theta _B^{\left( 2 \right)}$ for the second BS by applying similar procedures.
		
		Finally, the HRIS position ${{\mathbf{p}}_R}$can be estimated from the following least squares problem
		\begin{equation}\label{eq26}
			\scalebox{0.83}{$\displaystyle
				\begin{aligned}
					\mathop {\min }\limits_{{{\mathbf{p}}_R}} {\left( {{\mathbf{p}}_B^{\left( 1 \right)} - {{\mathbf{p}}_R}} \right)^{\text{T}}}{{\mathbf{G}}_1}\left( {{\mathbf{p}}_B^{\left( 1 \right)} - {{\mathbf{p}}_R}} \right) + {\left( {{\mathbf{p}}_B^{\left( 2 \right)} - {{\mathbf{p}}_R}} \right)^{\text{T}}}{{\mathbf{G}}_2}\left( {{\mathbf{p}}_B^{\left( 2 \right)} - {{\mathbf{p}}_R}} \right)
				\end{aligned}
				$}
		\end{equation}
		where ${{\mathbf{G}}_1} = {{\mathbf{I}}_3} - {\mathbf{g}}\left( {\varphi _B^{\left( 1 \right)},\theta _B^{\left( 1 \right)}} \right){{\mathbf{g}}^{\text{T}}}\left( {\varphi _B^{\left( 1 \right)},\theta _B^{\left( 1 \right)}} \right)$, and \({\mathbf{g}}\left( {\varphi _B^{\left( 1 \right)},\theta _B^{\left( 1 \right)}} \right) = \left[ \sin \left( {\varphi _B^{\left( 1 \right)}} \right)\cos \left( {\theta _B^{\left( 1 \right)}} \right),\sin \left( {\varphi _B^{\left( 1 \right)}} \right)\sin \left( {\theta _B^{\left( 1 \right)}} \right),\right.\)
		\(\left.\cos \left( {\varphi _B^{\left( 1 \right)}} \right) \right]\), and ${{\mathbf{G}}_2}$ has the same form as that of ${{\mathbf{G}}_1}$. Therefore, ${{\mathbf{p}}_R}$ can be easily computed as
		\begin{equation}\label{eq27}
			\begin{aligned}
				{{\mathbf{p}}_R} = {\left( {{{\mathbf{G}}_1} + {{\mathbf{G}}_2}} \right)^{ - 1}}\left( {{{\mathbf{G}}_1}{\mathbf{p}}_B^{\left( 1 \right)} + {{\mathbf{G}}_2}{\mathbf{p}}_B^{\left( 2 \right)}} \right)
			\end{aligned}
		\end{equation}
		
		The estimation of the HRIS location is summarized in Algorithm 3.
		\begin{table}[htbp]
			\centering
			\label{table3}
			\begin{tabular}{l p{0.43\textwidth}}
				\toprule
				\multicolumn{2}{l}{\textbf{Algorithm 3:} HRIS Location Estimation} \\
				\midrule
				\multicolumn{2}{l}{\textbf{input:} $\hat \varphi _k^{\mathbf{f}}$, $\hat \theta _k^{\mathbf{f}}$, $\hat r_k^{\mathbf{f}}$} \\
				\multicolumn{2}{l}{\textbf{output:} ${{\mathbf{p}}_R}$} \\
				1: & Construct ${\mathbf{\hat Y}}\left( t \right)$ according to (\ref{eq4}) with the estimate of $\hat \varphi _k^{\mathbf{f}}$, $\hat \theta _k^{\mathbf{f}}$ and $\hat r_k^{\mathbf{f}}$.\\
				2: & \textbf{for} $i \leftarrow 1$ \textbf{to} $2$ \textbf{do}\\
				3: & \quad Solve for ${\mathbf{T}}_{B,x}^{\left( i \right)}\left( {\mathbf{u}} \right)$ and ${\mathbf{T}}_{B,z}^{\left( i \right)}\left( {\mathbf{u}} \right)$ according to (\ref{eq25}) with const-\\
				& \quad ruction of ${{\mathbf{\hat Y}}_w}$. \\
				4: & \quad Apply the MUSIC algorithm to ${\mathbf{T}}_{B,x}^{\left( i \right)}\left( {\mathbf{u}} \right)$ and ${\mathbf{T}}_{B,z}^{\left( i \right)}\left( {\mathbf{u}} \right)$ to obtain \\
				& \quad $\phi _B^{\left( i \right)}$ and $\varphi _B^{\left( i \right)}$.\\
				5: & \quad Solve for $\theta _B^{\left( i \right)}$ using $\theta _B^{\left( i \right)} = \arccos \left( {\frac{{\phi _B^{\left( i \right)}}}{{\sin \left( {\varphi _B^{\left( i \right)}} \right)}}} \right)$.\\
				6:& \textbf{end}\\
				7:& Compute ${{\mathbf{p}}_R}$ according to (\ref{eq27}).\\
				\bottomrule
			\end{tabular}
		\end{table}
		
		\begin{remark}[Geometric Configuration and Observability]
			The triangulation-based HRIS localization in Algorithm 3 requires careful consideration of the BS geometric configuration to ensure position uniqueness. From (\ref{eq27}), the HRIS position is computed as a weighted average of the two BS positions, where the weighting matrices ${{\mathbf{G}}_1}$ and ${{\mathbf{G}}_2}$ project onto the subspaces orthogonal to the respective line-of-sight directions ${\mathbf{g}}\big( {\varphi _B^{( i )},\theta _B^{( i )}} \big)$. For ${{\mathbf{G}}_1}+{{\mathbf{G}}_2}$ to be invertible, the two direction vectors ${\mathbf{g}}\big( {\varphi _B^{( 1 )},\theta _B^{( 1 )}} \big)$ and ${\mathbf{g}}\big( {\varphi _B^{( 2 )},\theta _B^{( 2 )}} \big)$ must be linearly independent, which requires that the two BSs and the HRIS are not collinear.
			
			Moreover, the positioning accuracy is governed by the angular separation between the two BSs as viewed from the HRIS. When the BSs are nearly collinear with the HRIS, ${{\mathbf{G}}_1} + {{\mathbf{G}}_2}$ approaches a rank-deficient matrix, leading to amplified estimation errors perpendicular to the baseline connecting the two BSs. The optimal configuration occurs when the two BSs subtend a right angle at the HRIS, i.e., ${\mathbf{g}}\big( {\varphi _B^{\left( 1 \right)},\theta _B^{\left( 1 \right)}} \big) \perp {\mathbf{g}}\big( {\varphi _B^{\left( 2 \right)},\theta _B^{\left( 2 \right)}} \big)$, which minimizes the condition number of ${{\mathbf{G}}_1} + {{\mathbf{G}}_2}$ and balances the localization accuracy across all spatial dimensions. This geometric insight can guide the practical deployment of BSs in position-unknown HRIS scenarios.
		\end{remark}
		
		\subsection{HRIS Phase Shifts Optimization}
		\begin{figure}[htbp]
			\centering
			\centerline{\includegraphics[width=8cm]{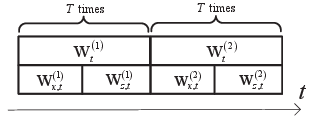}}
			\caption{Configuration for phase design.}
			\label{fig3}
		\end{figure}
		To improve the estimation accuracy of the HRIS position and, in turn, enhance the estimation accuracy of NF targets from the BS side, we sequentially optimize the HRIS phase matrix with the estimated $\hat \varphi _k^{\mathbf{f}}$, $\hat \theta _k^{\mathbf{f}}$, and $\hat r_k^{\mathbf{f}}$, as shown in Fig. \ref{fig3}.
		
		First, consider the HRIS phase optimization problem for the received signal of the first BS along the x-axis in the $t$-th time slot. To maximize the received signal power ${\left\| {\sqrt {1 - \delta } {\mathbf{H}}_x^{\left( 1 \right)}{\mathbf{W}}_{x,t}^{\left( 1 \right)}{\mathbf{\bar y}}\left( t \right)} \right\|^2}$ at the BS, subject to $\left| {{{\big[ {{\mathbf{W}}_{x,t}^{\left( 1 \right)}} \big]}_{i,i}}} \right| = 1,{\text{ }}i = 1, \cdots ,M$, the optimization problem can be cast as
		\begin{equation}\label{eq28}
			\begin{gathered}
				{\text{max }}{\left\| {\sqrt {1 - \delta } {\mathbf{H}}_x^{\left( 1 \right)}{\mathbf{W}}_{x,t}^{\left( 1 \right)}{\mathbf{\bar y}}\left( t \right)} \right\|^2} \hfill \\
				{\text{s}}{\text{.t}}{\text{. }}\left| {{{\big[ {{\mathbf{W}}_{x,t}^{\left( 1 \right)}} \big]}_{i,i}}} \right| = 1,{\text{ }}i = 1, \cdots ,M \hfill \\
			\end{gathered}
		\end{equation}
		
		Due to the unit circle constraint in (\ref{eq28}) being non-convex, solving it requires a non-convex method, leading to high complexity and the potential for a locally optimal solution. To avoid solving the non-convex problem (\ref{eq28}), realizing ${\mathbf{W}}_{x,t}^{\left( 1 \right)}$ is diagonal, we have $\sqrt {1 - \delta } {\mathbf{H}}_x^{\left( 1 \right)}{\mathbf{W}}_{x,t}^{\left( 1 \right)}{\mathbf{\bar y}}\left( t \right) = \sqrt {1 - \delta } {\mathbf{H}}_x^{\left( 1 \right)}{\text{diag}}\left( {{\mathbf{\bar y}}\left( t \right)} \right){\mathbf{w}}_{x,t}^{\left( 1 \right)}$.  Hence, (\ref{eq28}) can be rewritten as
		\begin{equation}\label{eq29}
			\begin{gathered}
				{\text{max }}\left( {{\mathbf{w}}_{x,t}^{\left( 1 \right)}} \right)_{}^{\text{H}}{\mathbf{Rw}}_{x,t}^{\left( 1 \right)} \hfill \\
				{\text{s}}{\text{.t}}{\text{. }}\left| {{{\big[ {{\mathbf{w}}_{x,t}^{\left( 1 \right)}} \big]}_i}} \right| = 1,{\text{ }}i = 1, \cdots ,M \hfill \\
			\end{gathered}
		\end{equation}
		where ${\mathbf{R}} = {{\mathbf{C}}^{\text{H}}}{\mathbf{C}}$, ${\mathbf{C}} = \sqrt {1 - \delta } {\mathbf{H}}_x^{\left( 1 \right)}{\text{diag}}\left( {{\mathbf{\bar y}}\left( t \right)} \right)$. It can be seen from (\ref{eq29}) that we have reformulated the non-convex optimization objective in (\ref{eq28}) into a quadratic optimization one. Further, by defining ${\mathbf{\tilde W}} = {\mathbf{w}}_{x,t}^{\left( 1 \right)}{\left( {{\mathbf{w}}_{x,t}^{\left( 1 \right)}} \right)^{\text{H}}}$, (\ref{eq29}) can be converted to
		\begin{equation}\label{eq30}
			\begin{gathered}
				{\text{max tr}}\big( {{\mathbf{R\tilde W}}} \big) \hfill \\
				{\text{s}}{\text{.t}}{\text{. }}{\mathbf{\tilde W}} \geqslant 0,{\text{ diag}}\left( {{\mathbf{\tilde W}}} \right) = {{\mathbf{1}}_{\text{M}}},{\text{ rank}}\left( {{\mathbf{\tilde W}}} \right) = 1 \hfill \\
			\end{gathered}
		\end{equation}
		
		By (\ref{eq30}), we have transformed the non-convex unit modulus constraint in (\ref{eq29}) into ${\text{diag}}\big( {{\mathbf{\tilde W}}} \big) = {{\mathbf{1}}_{\text{M}}}$, ${\mathbf{\tilde W}} \geqslant 0$, and ${\text{rank}}\big( {{\mathbf{\tilde W}}} \big) = 1$. Since ${\text{rank}}\big( {{\mathbf{\tilde W}}} \big) = 1$ is a non-convex constraint, we perform convex relaxation and reach an SDP-based convex relaxation problem
		\begin{equation}\label{eq31}
			\begin{gathered}
				{\text{max tr}}\big( {{\mathbf{R\tilde W}}} \big) \hfill \\
				{\text{s}}{\text{.t}}{\text{. }}{\mathbf{\tilde W}} \succeq 0,{\text{ diag}}\big( {{\mathbf{\tilde W}}} \big) = {{\mathbf{1}}_{\text{M}}} \hfill \\
			\end{gathered}
		\end{equation}
		
		With (\ref{eq31}), we can obtain the optimal solution ${{\mathbf{\tilde W}}^*}$ using a convex optimization toolbox such as CVX, but ${{\mathbf{\tilde W}}^*}$ may not satisfy the rank-one constraint. Therefore, to ensure the rank-one requirement, Gaussian randomization \cite{31} is adopted to find the $L$ suboptimal solutions ${\mathbf{w}}_{t,l}^*$ in (\ref{eq29}) from  ${{\mathbf{\tilde W}}^*}$. Let ${{\mathbf{\tilde W}}^*} = {\mathbf{UD}}{{\mathbf{U}}^{\text{H}}}$, ${\mathbf{w}}_{t,l}^*$ can be computed as
		\begin{equation}\label{eq32}
			\begin{aligned}
				{\mathbf{w}}_{t,l}^* = {\mathbf{U}}{{\mathbf{D}}^{\frac{1}{2}}}{{\mathbf{n}}_l},{\text{ }}l = 1, \cdots ,\mathcal{L}
			\end{aligned}
		\end{equation}
		where ${{\mathbf{n}}_l}$ is a Gaussian random vector with zero mean and variance $\sigma _n^2$.
		
		Finally, by maximizing ${\mathbf{w}}_t^{\text{H}}{\mathbf{R}}{{\mathbf{w}}_t}$, the optimal solution ${\mathbf{w}}_{t,l}^*$ can be selected from the $\mathcal{L}$ candidate solutions as the approximate optimal solution for (\ref{eq29}).  In other words, the final solution is
		\begin{equation}\label{eq33}
			\begin{aligned}
				{\mathbf{w}}_{t,{l^*}}^* = \mathop {\arg \max }\limits_{{\mathbf{w}}_{t,l}^{}} {\mathbf{w}}_{t,l}^{\text{H}}{\mathbf{R}}{{\mathbf{w}}_{t,l}},{\text{ }}l = 1, \cdots ,\mathcal{L}
			\end{aligned}
		\end{equation}
		
		The HRIS phase shift optimization method is summarized in Algorithm 4.
		\begin{table}[htbp]
			\centering
			\label{table4}
			\begin{tabular}{l p{0.43\textwidth}}
				\toprule
				\multicolumn{2}{l}{\textbf{Algorithm 4:} HRIS Phase Shift Optimization} \\
				\midrule
				\multicolumn{2}{l}{\textbf{input:} $\hat \varphi _k^{\mathbf{f}}$, $\hat \theta _k^{\mathbf{f}}$, $\hat r_k^{\mathbf{f}}$, ${\mathbf{H}}_x^{\left( 1 \right)}$, ${\mathbf{H}}_x^{\left( 2 \right)}$, ${\mathbf{H}}_z^{\left( 1 \right)}$, ${\mathbf{H}}_z^{\left( 2 \right)}$} \\
				\multicolumn{2}{l}{\textbf{output:} ${\mathbf{w}}_{t,{l^*}}^*$} \\
				1: & Construct ${\mathbf{\hat y}}\left( t \right)$ by estimating $\hat \varphi _k^{\mathbf{f}}$, $\hat \theta _k^{\mathbf{f}}$, $\hat r_k^{\mathbf{f}}$.\\
				2: & Obtain the optimal solution ${\mathbf{\tilde W}}$ using CVX for the SDP problem in (\ref{eq31}).\\
				3: & Perform eigenvalue decomposition on ${\mathbf{\tilde W}}$ to obtain ${\mathbf{U}}$ and ${\mathbf{D}}$.\\
				4: & \textbf{for} $l \leftarrow 1$ \textbf{to} $\mathcal{L}$ \textbf{do}\\
				5: & \quad Obtain the Gaussian randomization vector ${\mathbf{w}}_{t,l}^*$ using (\ref{eq32}).\\
				6: & \quad Calculate candidate vector ${\mathbf{w}}_{t,{l^*}}^*$ using (\ref{eq33}).\\
				7: & \textbf{end}\\
				\bottomrule
			\end{tabular}
		\end{table}
		
		The same procedure of HRIS phase shift optimization is applied to the received signal of the first BS along the z-axis, as well as that for the second BS.
		
		The proposed two-stage framework systematically addresses the four challenges outlined in Section II: (1) The DANM algorithm decouples the NF angle-range coupling by transforming the NF model into a virtual FF model through delay-based cross-correlation, enabling separate estimation of 2-D angles and range via MUSIC. (2) The TLS correction method mitigates the systematic errors introduced by the Fresnel approximation by using the exact NF model. (3) The ANM-based angle estimation combined with the LS geometric triangulation determines the unknown HRIS position ${{\mathbf{p}}_R}$ from two BS observations, enabling coordinate transformation from HRIS-centric to BS-centric reference frames. (4) The SDR-based phase optimization enhances HRIS self-localization accuracy, which in turn improves NF target positioning in the absolute coordinates. Having established the complete algorithmic framework, we now turn to deriving theoretical performance bounds to benchmark the achievable accuracy limits.
		
		\section{Performance Bound Analysis}
		To evaluate the theoretical performance limits of the proposed position-unknown HRIS-aided localization framework, this section derives the Fisher information matrix (FIM) and the corresponding Cramér-Rao bound (CRB) for NF target parameters, as well as the position error bound (PEB) for the HRIS position. From a signal processing perspective, the CRB characterizes the minimum achievable variance for any unbiased estimator and serves as a fundamental lower bound dictated by the observation model structure, noise statistics, and number of snapshots. However, it is crucial to recognize that the CRB is only achievable when the estimator is unbiased, in the presence of model mismatch (such as the Fresnel approximation in our NF model). Systematic bias prevents the estimator from approaching the CRB even at high SNR, resulting in an error floor. This is why the proposed TLS correction, which mitigates the mismatch-induced bias, is essential for enabling near-CRB performance in angule estimation.
		
		Moreover, the phase optimization in Challenge 4 directly impacts the FIM, and hence, the CRB and PEB. Specifically, optimizing ${\mathbf{W}}_{x,t}^{\left( i \right)}$ to maximize received signal power at a BS effectively increases the signal-to-noise ratio (SNR) of the HRIS-to-BS channel, which enlarges the diagonal entries of the FIM ${\mathbf{F}}_B^{}\left( {\bm{\mu}} \right)$ (\ref{eq36}). Since the CRB is the inverse of the FIM, increasing the FIM reduces the CRB, thereby reducing the theoretical lower bound and improving the achievable estimation accuracy. This information-theoretic interpretation explains why phase optimization not only enhances empirical performance, but also fundamentally improves the best possible performance as quantified by the bounds.
		
		Let the unknown parameter vector be ${{\bm{\mu }}_u} = {\left[ {{\bm{\mu }}_0^{\text{T}}, \cdots ,{\bm{\mu }}_k^{\text{T}}, \cdots ,{\bm{\mu }}_K^{\text{T}}} \right]^{\text{T}}} \in {\mathbb{R}^{3K}}$, where ${\bm{\mu }}_k^{} = {\left[ {{\varphi _k},{\theta _k},{r_k}} \right]^{\text{T}}}$, and ${\bm{\mu }}_B^{\left( 1 \right)} = {\big[ {\theta _B^{\left( 1 \right)},\varphi _B^{\left( 1 \right)}} \big]^{\text{T}}}$, ${\bm{\mu }}_B^{\left( 2 \right)} = {\big[ {\theta _B^{\left( 2 \right)},\varphi _B^{\left( 2 \right)}} \big]^{\text{T}}}$. The Fisher information matrix ${{\mathbf{F}}_u}\left( {\bm{\mu}}_u \right)$ with respect to $ {{\bm{\mu }}_u}$ can be expressed as \cite{32}
		\begin{equation}\label{eq34}
			\begin{aligned}
				{{\mathbf{F}}_u}\left( {\bm{\mu}}_u \right) = \frac{2}{{{\sigma ^2}}}\sum\limits_{t = 1}^T {{\text{Re}}\left\{ {{{\left( {\frac{{\partial {\mathbf{\bar y}}\left( t \right)}}{{\partial {\bm{\mu}}}_u}} \right)}^{\text{H}}}\frac{{\partial {\mathbf{\bar y}}\left( t \right)}}{{\partial {\bm{\mu}}}_u}} \right\}}
			\end{aligned}
		\end{equation}
		where the detailed derivation of the Fisher information matrix for each parameter set of an NF target can be found in Appendix A, and the corresponding CRB can be obtained from ${{\mathbf{F}}_u}{\left( {\bm{\mu}_u} \right)^{ - 1}} \in {\mathbb{R}^{3K \times 3K}}$.
		
		For the HRIS, the PEB can be obtained as \cite{33}
		\begin{equation}\label{eq35}
			\begin{aligned}
				{\text{PEB = }}\sqrt {{\text{tr}}\big( {{\mathbf{F}}{{\left( {\mathbf{p}} \right)}^{ - 1}}} \big)}
			\end{aligned}
		\end{equation}
		where ${\mathbf{F}}\big( {\mathbf{p}} \big) = {\mathbf{JF}}_B^{}\big( {{\bm{\mu}}_B^{\left( 1 \right)}} \big){{\mathbf{J}}^{\text{T}}} + {\mathbf{JF}}_B^{}\big( {{\bm{\mu}}_B^{\left( 2 \right)}} \big){{\mathbf{J}}^{\text{T}}}$, ${\mathbf{J}} \in {\mathbb{R}^{3 \times 2}}$ represents the Jacobian matrix of ${\mathbf{J}} = \left[ {\frac{{\partial \theta _B^{\left( 1 \right)}}}{{\partial {\mathbf{p}}_R^{}}},\frac{{\partial \varphi _B^{\left( 1 \right)}}}{{\partial {\mathbf{p}}_R^{}}}} \right]$, and ${\mathbf{F}}_B^{}\left( {\bm{\mu}}_B \right)$ can be represented as
		\begin{equation}\label{eq36}
			\begin{aligned}
				{\mathbf{F}}_B^{}\left( {\bm{\mu}}_B \right) = \frac{2}{{{\sigma ^2}}}\sum\limits_{t = 1}^T {{\text{Re}}\left\{ {{{\left( {\frac{{\partial {\mathbf{\tilde y}}\left( t \right)}}{{\partial {\bm{\mu}}}_B}} \right)}^{\text{H}}}\frac{{\partial {\mathbf{\tilde y}}\left( t \right)}}{{\partial {\bm{\mu}}}_B}} \right\}}
			\end{aligned}
		\end{equation}
		where ${\mathbf{\tilde y}}\left( t \right) = \sqrt {1 - \delta } \left[ {\begin{array}{*{20}{c}}
				{{{\mathbf{H}}_x}} \\
				{{{\mathbf{H}}_z}}
		\end{array}} \right]{{\mathbf{W}}_t}{\mathbf{\bar y}}\left( t \right)$. The detailed derivations of the PEB for the HRIS location parameters are provided in Appendix B.
		
		\begin{corollary}[Asymptotic Scaling of CRB and PEB]
			The CRB and PEB exhibit predictable asymptotic behavior as functions of system parameters. Specifically:
			\begin{itemize}
				\item \textbf{SNR scaling:} For high SNR, the CRB for the NF target parameters scales as ${{\mathbf{F}}_u}{\left( {\bm{\mu}_u} \right)^{ - 1}} \propto {\sigma ^2}/T$ from (\ref{eq34}), implying that the angle and range RMSEs decrease proportionally to $1/\sqrt {{\text{SNR}} \cdot T} $ in the unbiased regime. However, under model mismatch (Fresnel approximation), the RMSE converges to a bias-dominated floor independent of SNR, as ${\text{RMSE}} \to \sqrt {{\text{bias}}^2 + {\text{CRB}}} \approx {\text{bias}}$ for large SNR.
				\item \textbf{Array aperture scaling:} The Fisher information for the angle parameters scales as ${\mathbf{F}}_{\varphi\varphi} \propto {d^2} \cdot \left( {\sum\limits_{m =  - {m_x}}^{{m_x}} {{m^2}}  + \sum\limits_{n =  - {n_z}}^{{n_z}} {{n^2}} } \right) \propto {d^2} \cdot \left( {m_x^3{n_z} + {m_x}n_z^3} \right)$ where $M = \left( {2{m_x} + 1} \right) \times \left( {2{n_z} + 1} \right)$. For a square array with ${m_x} = {n_z} \approx \sqrt {M/4} $, it simplifies to ${\mathbf{F}}_{\varphi\varphi} \propto {d^2}{M^{3/2}}$, giving ${\text{CRB}} \propto 1/\left( {{d^2}{M^{3/2}}} \right)$. Thus, doubling the array size along both dimensions (quadrupling $M$) reduces CRB by a factor of 8, while doubling element spacing $d$ reduces CRB by a factor of 4, highlighting the strong dependence on the effective aperture $D = d\sqrt M $.
				\item \textbf{PEB dependence on Jacobian conditioning:} From (\ref{eq35}), the PEB depends on the condition number of the Jacobian matrix ${\mathbf{J}}$, which encodes the sensitivity of HRIS-to-BS angles to HRIS position perturbations. When the two BSs subtend a small angle at the HRIS (near-collinear configuration), ${\mathbf{J}}$ becomes ill-conditioned, causing the PEB to grow unboundedly even if the angle FIM ${{\mathbf{F}}_B}$ remains well-conditioned.
			\end{itemize}
			These asymptotic insights guide system design: increasing array size and SNR improves accuracy only when bias is mitigated (via TLS), and BS placement should maximize angular diversity to minimize PEB.
		\end{corollary}
		
		The derived CRB for the NF target parameters $\left\{ {{\varphi _k},{\theta _k},{r_k}} \right\}$ and PEB for the HRIS position ${{\mathbf{p}}_R}$ provide theoretical lower bounds on the estimation variance for any unbiased estimator under the given system configuration and noise level. These bounds are functions of system parameters, including the number of HRIS elements, BS array configurations, snapshot count, and SNR. In practice, the actual performance of the proposed DANM, MUSIC, TLS correction, ANM, and LS triangulation algorithms depends on how effectively they exploit the available observations. The gap between algorithmic performance and theoretical bounds reveals the potential for further refinement. To comprehensively evaluate the practical performance of the proposed position-unknown HRIS-aided localization framework and validate its proximity to the theoretical limits, we now present extensive simulation results across various scenarios.
		
		\section{Simulation Results}
		This section validates the proposed two-stage gridless localization framework through comprehensive Monte Carlo simulations. The experiments are designed to: (1) evaluate the NF target parameter estimation accuracy of the DANM and MUSIC algorithms versus the theoretical CRB, (2) demonstrate the effectiveness of TLS correction in mitigating the Fresnel approximation errors, (3) assess the HRIS self-localization accuracy via the ANM and LS triangulation versus the theoretical PEB, and (4) quantify the performance enhancement achieved by the SDR-based phase optimization. In all experiments, the array element spacing is $d = \lambda /4$, and the total number of snapshots is $T$. The definition of SNR is
		\begin{equation}\label{eq37}
			\begin{aligned}
				{\text{SNR = 10lo}}{{\text{g}}_{10}}\frac{{\sum\limits_{m =  - {m_x}}^{{m_x}} {\sum\limits_{n =  - {n_z}}^{{n_z}} {\sum\limits_{t = 1}^T {{{\left| {y_{m,n}^{}\left( t \right)} \right|}^2}} } } }}{{\sigma _n^2MT}}
			\end{aligned}
		\end{equation}
		
		The root mean square error (RMSE) for parameter estimation is obtained by $RMSE = \sqrt {\frac{1}{{\tilde MK}}\sum\limits_{\tilde m = 1}^{\tilde M} {\sum\limits_{k = 1}^K {{{\left( {{{\hat \eta }_{\tilde m,k}} - {\eta _k}} \right)}^2}} } } $, where ${\hat \eta _{\tilde m,k}}$ denotes the estimated value of the $k$-th source parameter in the $\tilde m$-th Monte Carlo experiment, $\tilde M$ is the number of Monte Carlo experiments, ${\eta _k}$ is the corresponding theoretical value, and $\eta $ represents the azimuth $\theta $, elevation $\varphi $, or range $r$.
		
		\subsection{Experiment 1: NF Target Estimation Performance}
		\begin{figure}[!t]
			\centering
			\centerline{\includegraphics[width=0.9\linewidth]{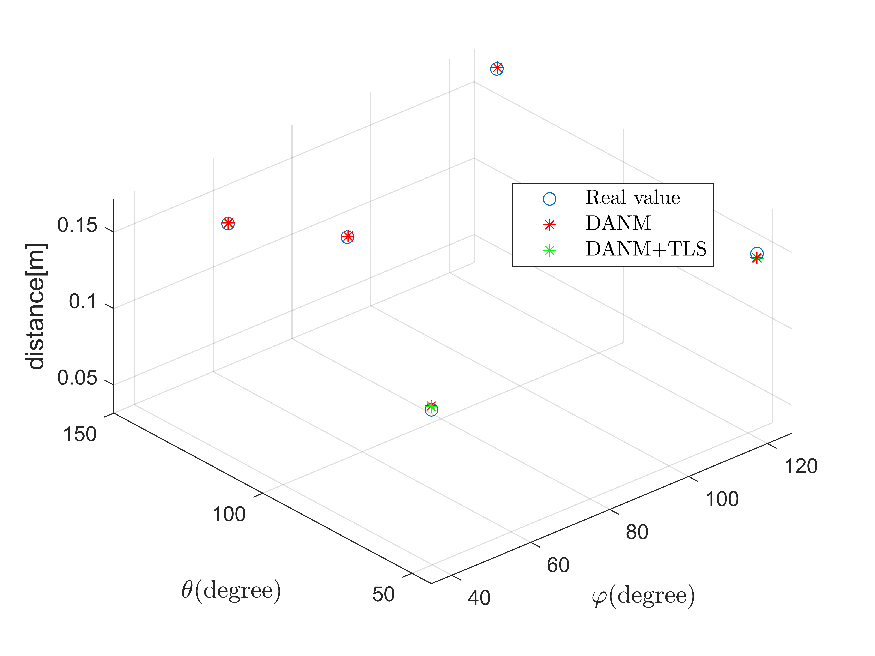}}
			\caption{3D Scatter Plot of NF Signal Sources.}
			\label{san}
		\end{figure}
		\begin{table}[!t]
			\centering
			\setlength{\tabcolsep}{10pt}
			\caption{{{RMSEs} of estimated parameters for {the} five NF sources (SNR = 25 dB).}}
			\label{tab:source1}
			\begin{tabular}{@{}lccc@{}}
				\toprule
				method & \(\varphi\) {(deg.)} & \(\theta\) {(deg.)} & \(r\) {(m)} \\
				\midrule
				DANM     & \(8.17\times10^{-2}\) & \( 2.443\times10^{-1}\) & \(1.877\times10^{-3}\) \\
				DANM+TLS   & \(1.79\times10^{-2}\) & \(1.333\times10^{-1}\) & \(1.853\times10^{-3}\) \\
				\bottomrule
			\end{tabular}
		\end{table}
		In this experiment, five narrowband NF sources are positioned at $\left( 34.8194^{\circ},\ 43.4387^{\circ},\ 0.1454\text{m} \right)$, $\left( {{{45.2282}^\circ },{{125.4274}^\circ },0.1701{\text{m}}} \right)$, \(\left( {{77.5605}^\circ },{{128.1999}^\circ },0.12 \right.\)
		\(\left.35{\text{m}}\right)\), $\left( {{{124.8440}^\circ },{{140.6266}^\circ },0.1696{\text{m}}} \right)$, \(\left( {{126.1008}^\circ },55.\right.\)
		\(\left.{0334}^\circ ,0.1371{\text{m}} \right)\), impinging on HRIS shown in Fig. \ref{fig1}. The SNR is 25 dB, and the number of snapshots is 500, including 125 pseudo snapshots. The 3-D scatter plot of the five NF source positions relative to the HRIS is shown in Fig. \ref{san}. {The {corresponding RMSEs} of the five NF source {position} parameters {are listed} in Table \ref{tab:source1}.} The {results reveal} several critical insights: (1) The DANM+TLS algorithm (green circles) achieves significantly higher localization accuracy compared to standard DANM (red stars), with estimated positions closely clustering around the true target locations (blue circles). (2) The systematic error compensation provided by TLS effectively mitigates the bias introduced by the Fresnel approximation, particularly evident in the reduced scatter of DANM+TLS estimates for the angle parameters. (3) All five targets are successfully detected and their 2-D angles are correctly paired with corresponding range estimates, demonstrating the robustness of the proposed pairing algorithm even in challenging multi-target scenarios. (4) Range estimation shows relatively larger scatter for both methods compared to angle estimation, which is consistent with the physics of the problem: after fixing the linear phase terms from angle estimation, range information must be extracted from much weaker quadratic phase variations $\phi _{}^k = \frac{{{d^2}}}{{2r_{0,0}^k}}$, making it inherently more susceptible to noise contamination.
		\subsection{Experiment 2: NF Source Parameters' RMSE versus SNR}
		This experiment evaluates the robustness of both algorithms against varying noise levels using two narrowband NF sources at $\left( {{{48.7223}^\circ },{{138.5400}^\circ },0.1503\,{\text{m}}} \right)$ and $\left( {{{124.8461}^\circ },{{140.6148}^\circ },0.1696\,{\text{m}}} \right)$, with a 17$ \times $ 17 HRIS array (${m_x} = {n_z} = 9$). The SNR varies from -10 dB to 20 dB and the number of snapshots is 500 (including 125 pseudo-snapshots).
		\begin{figure*}[htbp]
			\centering
			\subfigure[RMSE-$\varphi $]{
				\includegraphics[width=0.31\linewidth]{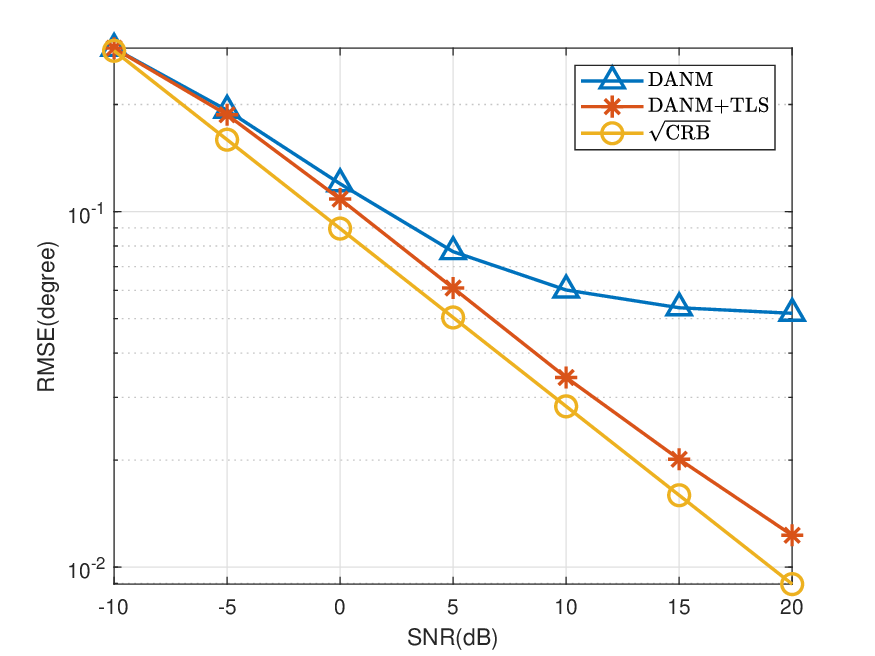}
			}
			\hfill
			\subfigure[RMSE-$\theta $]{
				\includegraphics[width=0.31\linewidth]{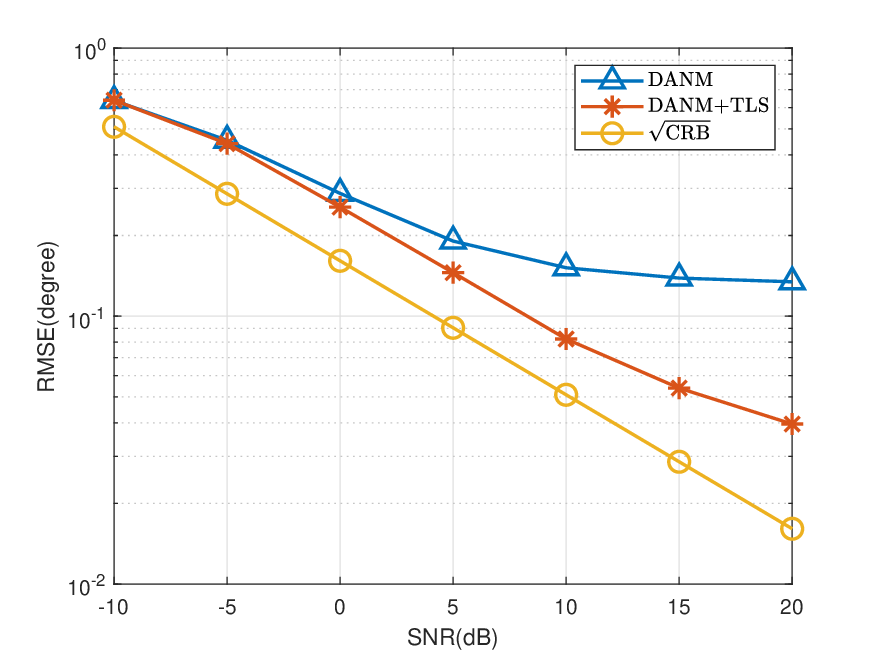}
			}
			\hfill
			\subfigure[RMSE-$r$]{
				\includegraphics[width=0.31\linewidth]{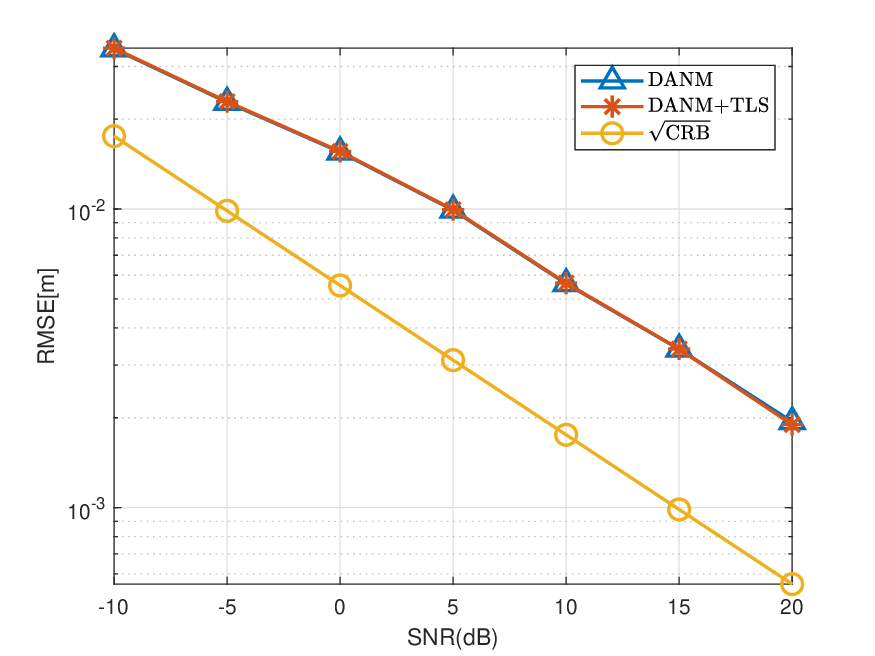}
			}
			\caption{NF Signal Source Parameters RMSE versus SNR.}
			\label{fig5}
		\end{figure*}
		Fig. \ref{fig5} reveals important performance characteristics of the proposed DANM+TLS method compared to the standard DANM across different SNR conditions. For angle estimation (both azimuth $\varphi$ and elevation $\theta$), DANM+TLS consistently outperforms DANM across all SNR levels, with the performance gap widening at higher SNRs. This trend occurs because at low SNRs, measurement noise dominates the estimation error, masking the benefits of systematic bias correction. However, as noise reduces with increasing SNR, the Fresnel approximation bias becomes the primary limiting factor, making TLS correction increasingly valuable. In contrast, range estimation shows only marginal improvement with TLS correction due to the sequential estimation strategy employed. Since angles are estimated first from the dominant linear phase terms (proportional to element indices $m$ and $n$), range estimation subsequently relies solely on the weaker quadratic phase terms ($\phi _x^k{m^2}$ and $\phi _z^k{n^2}$). These quadratic coefficients $\phi _x^k = \frac{{{d^2}}}{{2r_{0,0}^k}}\left( {1 - {{\sin }^2}{\varphi _k}{{\cos }^2}{\theta _k}} \right)$ produce much smaller phase variations compared to the linear terms, making range estimation inherently more noise-sensitive and less affected by systematic bias correction. All algorithms demonstrate logarithmic RMSE reduction with increasing SNR, consistent with CRB theory, but DANM+TLS approaches a lower error floor across all parameters, indicating superior bias mitigation capability.
		
		\subsection{Experiment 3: NF Source Parameters RMSE versus Snapshots}
		This experiment analyzes the impact of temporal diversity on estimation performance using the same two-target scenario as Experiment 2. With the SNR fixed at 5 dB, the total number of snapshots varies from 100 to 1200, and we maintain the number of pseudo-snapshots at 25\% of the total to preserve the cross-correlation structure.
		\begin{figure*}[htbp]
			\centering
			\subfigure[RMSE-$\varphi $]{
				\includegraphics[width=0.31\linewidth]{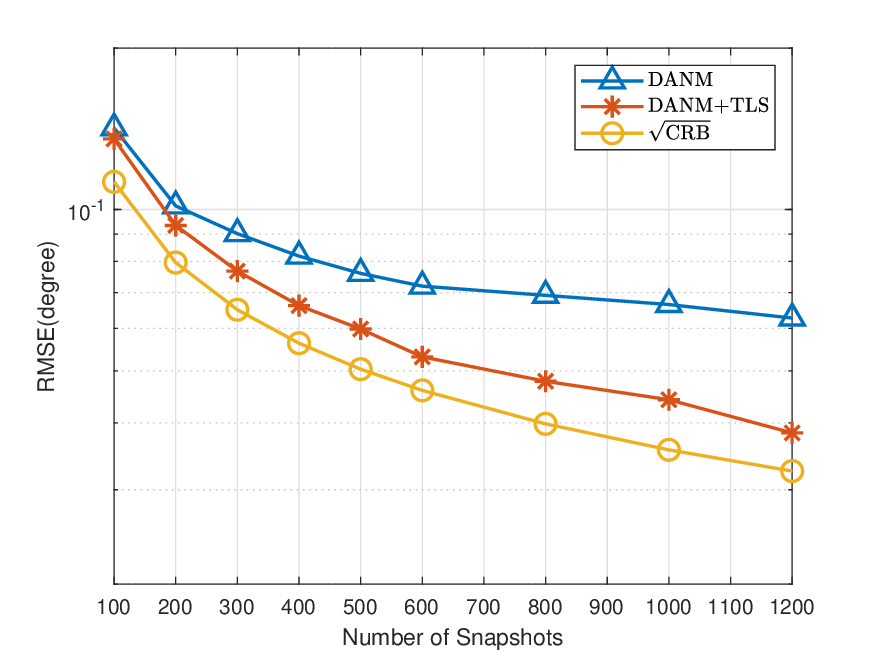}
			}
			\hfill
			\subfigure[RMSE-$\theta$]{
				\includegraphics[width=0.31\linewidth]{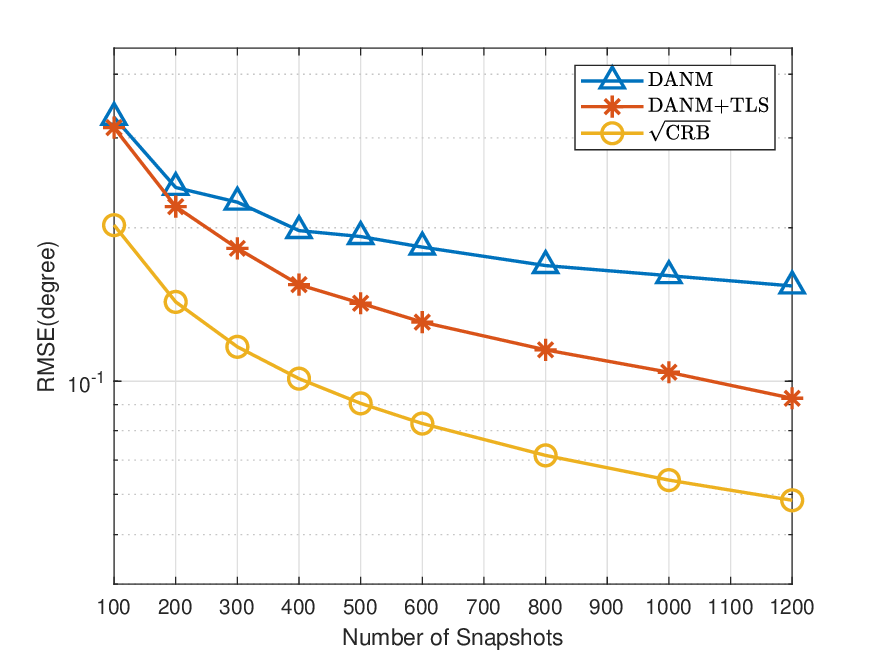}
			}
			\hfill
			\subfigure[RMSE-$r$]{
				\includegraphics[width=0.31\linewidth]{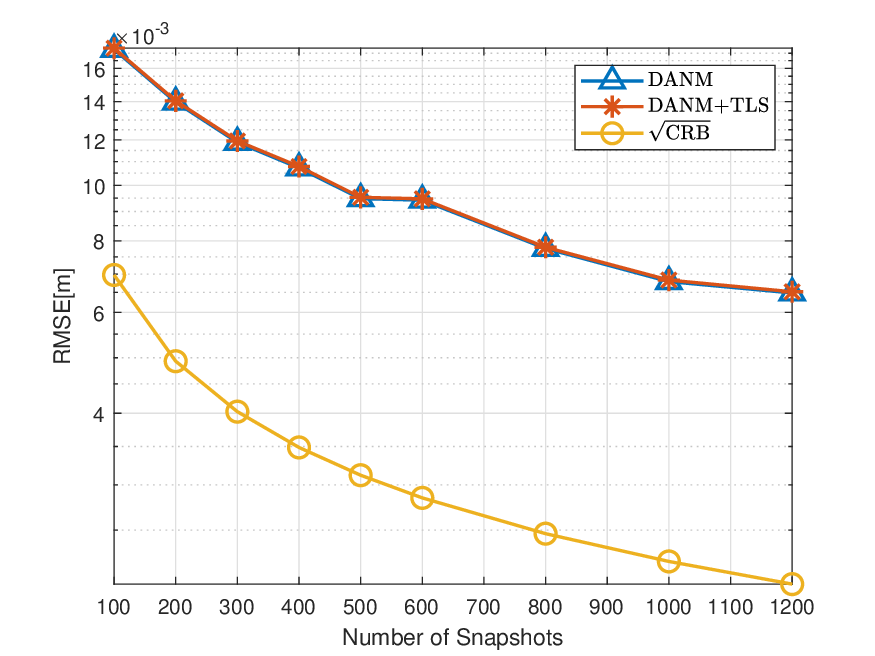}
			}
			\caption{NF Signal Source Parameters RMSE versus Snapshot.}
			\label{fig6}
		\end{figure*}
		
		Fig. \ref{fig6} demonstrates important convergence characteristics as a function of snapshot count. DANM+TLS achieves superior angle estimation accuracy compared to standard DANM across all snapshot numbers. This behavior indicates that the TLS bias correction provides consistent benefits regardless of the temporal averaging effects. Both algorithms exhibit the expected $1/\sqrt N $ RMSE reduction with increasing snapshots $N$, consistent with central limit theorem behavior. However, the performance curves reveal a fundamental difference: DANM converges to a biased estimate (higher error floor), while DANM+TLS converges toward the true parameter values due to systematic bias correction. The persistent gap between the two methods indicates that systematic errors in DANM cannot be eliminated by simply increasing sample size. Similar to Experiment 2, the range RMSE shows minimal difference between the two methods because range information is extracted from the quadratic phase terms after angle estimation, where the phase sensitivity $\phi _{}^k = \frac{{{d^2}}}{{2r_{0,0}^k}}$ becomes progressively weaker at longer distances, making it inherently noise-dominated rather than bias-limited in the considered SNR range.
		
		\subsection{Experiment 4: Performance of HRIS positioning and phase shift optimization} In this experiment, the number of HRIS elements and target position parameters are the same as in Experiment 2. The two BSs have an identical cross-shaped array, with ${N_x} = {N_z} = 5$. They are positioned at $\left( {{{40}^\circ },{{72}^\circ },10\,{\text{m}}} \right)$ and $\left( {{{30}^\circ },{{72}^\circ },30\,{\text{m}}} \right)$. Keeping 500 snapshots (including 125 pseudo snapshots), the SNR varies from -10 dB to 20 dB.
		\begin{figure*}[htbp]
			\centering
			\subfigure[RMSE-${\varphi _B} $]{
				\includegraphics[width=0.48\linewidth]{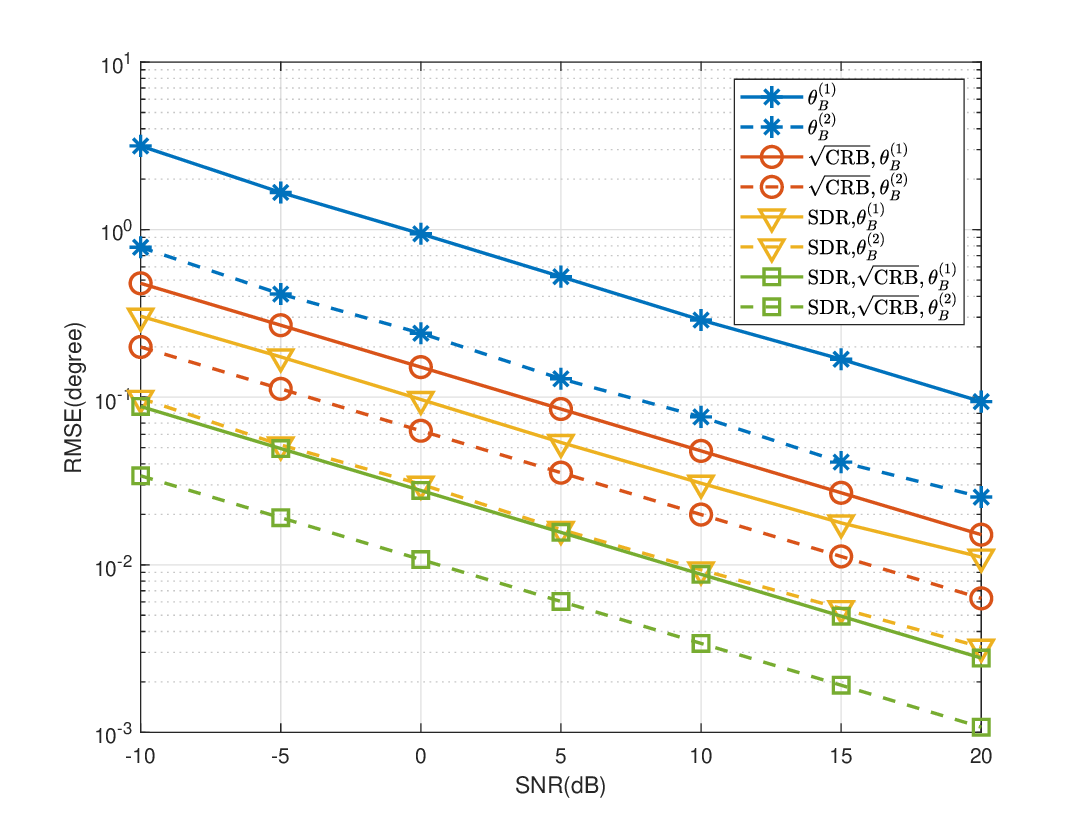}
			}
			\hfill
			\subfigure[RMSE-${\theta _B}$]{
				\includegraphics[width=0.48\linewidth]{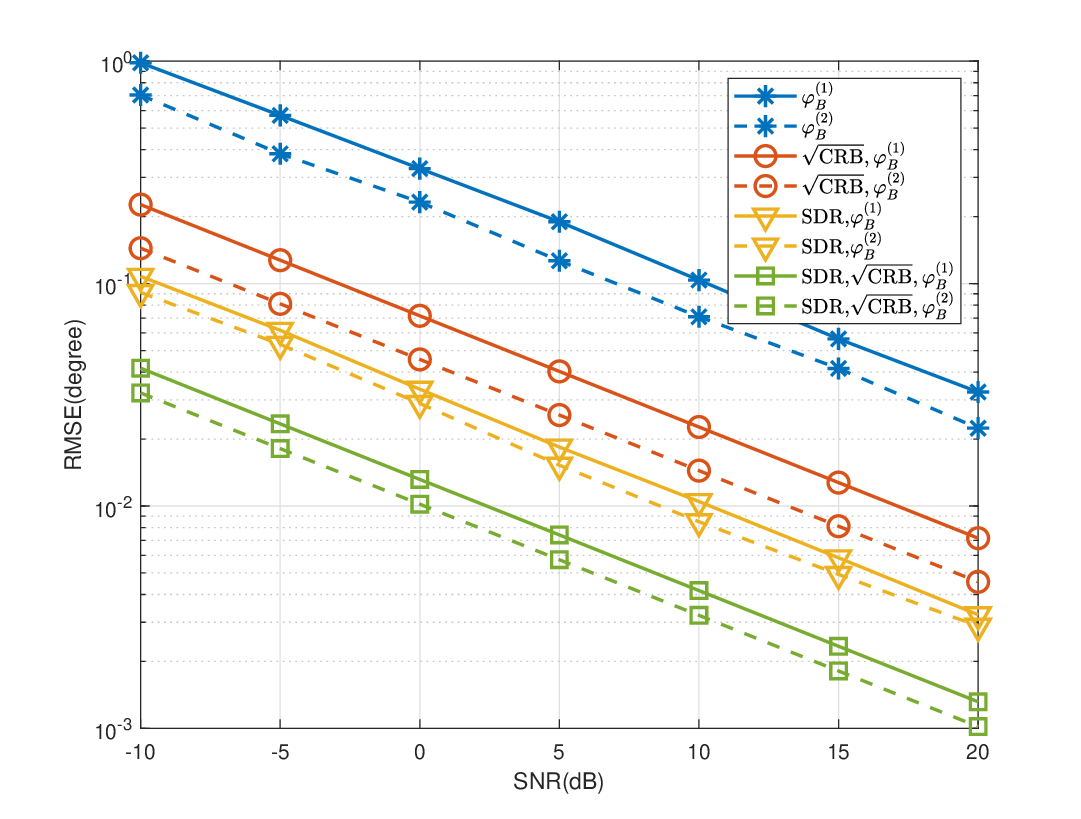}
			}
			\caption{HRIS angle parameters' RMSE versus SNR. }
			\label{fig7}
		\end{figure*}

		\begin{figure}[!t]
			\centering
			\centerline{\includegraphics[width=0.9\linewidth]{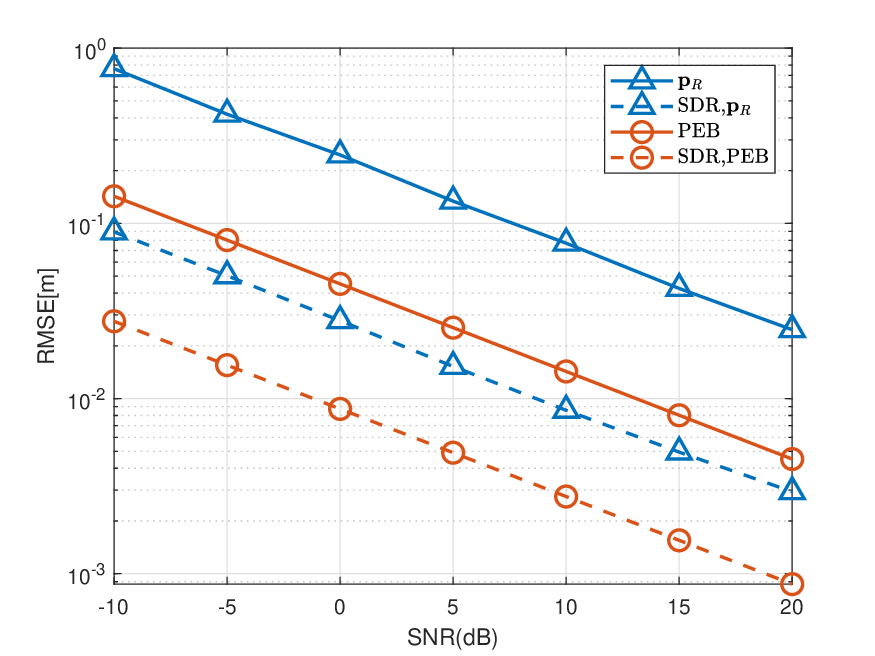}}
			\caption{HRIS position parameters RMSE versus SNR}
			\label{fig8}
		\end{figure}

		\begin{figure}[!t]
			\centering
			\centerline{\includegraphics[width=0.9\linewidth]{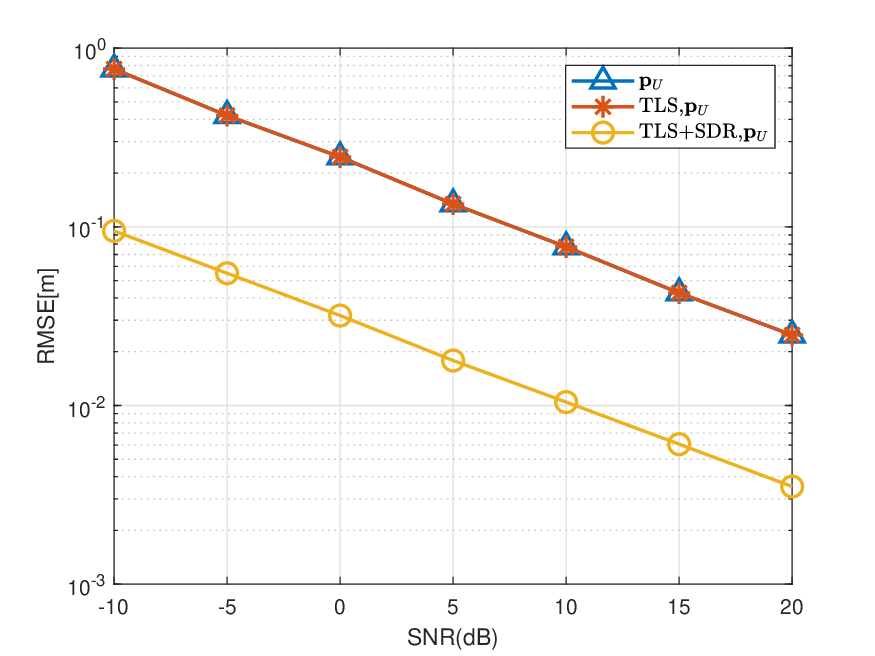}}
			\caption{UE position parameters RMSE versus SNR.}
			\label{fig9}
		\end{figure}
		
		Figs. \ref{fig7}, \ref{fig8}, and \ref{fig9} collectively demonstrate the effectiveness of the SDR-based HRIS phase optimization in enhancing the overall localization framework. {In the legends,} {``$\sqrt {{\text{CRB}}},\theta _B^{}$'' and ``$\sqrt {{\text{CRB}}} ,\varphi _B^{}$'' denote the corresponding square roots of { the CRBs} without phase optimization. The curves prefixed with ${\text{SDR}}$ indicate results obtained after phase optimization using SDR.  {Specifically,} ``${\text{SDR}},\theta _B^{}$'' and ``${\text{SDR}},\varphi _B^{}$'' are the RMSE after phase optimization{,} ``${\text{SDR}},\sqrt {{\text{CRB}}},\theta _B^{}$'' and ``${\text{SDR}},\sqrt {{\text{CRB}}} ,\varphi _B^{}$'' are the corresponding CRB square roots under the optimized phase configuration{, and} ``${\text{SDR}}, {{\text{PEB}}}$'' represents the corresponding {\text{PEB}} under the optimized phase configuration. Additionally, the curve labeled ``$\text{TLS + SDR},{{\mathbf{p}}_U}$'' shows the RMSE of the absolute position of NF targets using the combined {\text{TLS}} and {\text{SDR}} approach.} Fig. \ref{fig7} shows that after phase optimization, the estimation accuracy of the HRIS angles ${\varphi _B}$ and ${\theta _B}$ is consistently better than that of the non-optimized case across all SNR conditions. This improvement occurs because the SDR-based phase optimization enhances the received signal power carrying the HRIS angle information at the BS, thereby effectively improving angle estimation accuracy. Notably, the RMSE of estimated angles after phase optimization approaches to the corresponding CRB more closely under the same optimized phase configuration, indicating that phase optimization enables the system to achieve near-theoretical optimal performance. The CRBs themselves also decrease after optimization, reflecting the fundamental improvement in estimation conditions.
		
		Fig. \ref{fig8} reveals that the improved HRIS angle estimation directly translates to enhanced HRIS position estimation performance. Across all SNR conditions, phase optimization using SDR consistently reduces HRIS position parameter RMSE compared to the non-optimized case, with the corresponding PEB showing similar trends. It demonstrates that SDR-based phase optimization benefits not only the angle estimation of HRIS relative to BS but also the subsequent geometric triangulation for HRIS position determination.
		
		Finally, Fig. \ref{fig9} illustrates the cascading effect on NF target positioning. Since the UE global coordinates rely on the HRIS position through coordinate transformation, improving the estimation accuracy of HRIS angles ${\varphi _B}$ and ${\theta _B}$ via phase optimization effectively enhances the overall UE positioning accuracy, validating the closed-loop enhancement mechanism of the proposed two-stage framework.
		
		\section{Conclusion}
		This paper addresses the fundamental challenge of NF target localization when the RIS position is unknown, which alters the parameter identifiability conditions in RIS-aided sensing. We propose a two-stage gridless framework leveraging the dual sensing-reflecting capability of HRIS to systematically address NF parameter coupling, HRIS self-localization, model mismatch correction, and phase optimization. The key innovations include: (1) the delay-based virtual FF transformation enabling gridless DANM for decoupled angle-range estimation; (2) the TLS correction mitigating Fresnel approximation bias to achieve near-CRB performance at high SNR; (3) the ANM-based geometric triangulation for HRIS position determination; and (4) the SDR-based phase optimization maximizing the Fisher information to reduce both CRB and PEB. Extensive simulations validate that the proposed algorithms approach the derived theoretical bound performance across the SNR and snapshot regimes, with the bias-corrected method avoiding mismatch-induced error floors.
		
		\appendix
		\subsection{Calculate ${{\mathbf{F}}_u}\left( {\mathbf{\mu}_u} \right)$}
		In this appendix, the $\frac{{\partial {\mathbf{\bar y}}\left( t \right)}}{{\partial {\bm{\mu}}_u}}$ of ${{\mathbf{F}}_u}\left( {\bm{\mu}_u} \right)$ in (\ref{eq34}) for each parameter of an NF target is derived.   As $r_{m,n}^k = \sqrt {{{\left( {r_{0,0}^k} \right)}^2} - 2dr_{0,0}^k{\omega _k} + \left( {{m^2} + {n^2}} \right){d^2}} $, where ${\omega _k} = m\sin {\varphi _k}\cos {\theta _k} + n\cos {\varphi _k}$, and $r_k = r^k_{0,0}$, we have
		\begin{equation}\label{eq38}
			\scalebox{0.8}{$\displaystyle
				\begin{aligned}
					{\left[ {\frac{{\partial {\mathbf{Y}}\left( t \right)}}{{\partial {\varphi _k}}}} \right]_{m,n}}{\text{ = }}\frac{{j2\pi {r_k}d}}{{\lambda r_{m,n}^k}}\sqrt \delta  {e^{j\frac{{2\pi }}{\lambda }\left( {r_{m,n}^k - {r_k}} \right)}}\left( {n\sin {\varphi _k} - m\cos {\varphi _k}\cos {\theta _k}} \right)s_k^{}\left( t \right)
				\end{aligned}
				$}
		\end{equation}
		\begin{equation}\label{eq39}
			\scalebox{0.9}{$\displaystyle
				\begin{aligned}
					{\left[ {\frac{{\partial {\mathbf{Y}}\left( t \right)}}{{\partial {\theta _k}}}} \right]_{m,n}}{\text{ = }}\frac{{j2\pi {r_k}d}}{{\lambda r_{m,n}^k}}\sqrt \delta  {e^{j\frac{{2\pi }}{\lambda }\left( {r_{m,n}^k - {r_k}} \right)}}\left( {m\sin {\varphi _k}\sin {\theta _k}} \right)s_k^{}\left( t \right)
				\end{aligned}
				$}
		\end{equation}
		\begin{equation}\label{eq40}
			\scalebox{0.9}{$\displaystyle
				\begin{aligned}
					{\left[ {\frac{{\partial {\mathbf{Y}}\left( t \right)}}{{\partial {r_k}}}} \right]_{m,n}}{\text{  = }}\frac{{j2\pi }}{\lambda }\sqrt \delta  {e^{j\frac{{2\pi }}{\lambda }\left( {r_{m,n}^k - {r_k}} \right)}}\left( {\frac{{{r_k} - {\omega _k}d}}{{r_{m,n}^k}} - 1} \right)s_k^{}\left( t \right)
				\end{aligned}
				$}
		\end{equation}
		
		According to (\ref{eq38})-(\ref{eq40}), the gradient matrices ${{\mathbf{Y'}}_{{\varphi _k}}}\left( t \right)$, ${{\mathbf{Y'}}_{{\theta _k}}}\left( t \right)$, and ${{\mathbf{Y'}}_{{r_k}}}\left( t \right)$ of the overall HRIS received signal with respect to ${\varphi _k}$, ${\theta _k}$, and ${r_k}$, respectively, are derived by following the same arrangement rule for $m$ and $n$ as in (\ref{eq4}). As $\frac{{\partial {\mathbf{\bar y}}\left( t \right)}}{{\partial {\bm{\mu}_u}}} = {\text{vec}}\left( {\frac{{\partial {\mathbf{Y}}\left( t \right)}}{{\partial {\bm{\mu}}_u}}} \right)$, hence, $\frac{{\partial {\mathbf{\bar y}}\left( t \right)}}{{\partial {\varphi _k}}} = {\text{vec}}\left( {{{{\mathbf{Y'}}}_{{\varphi _k}}}\left( t \right)} \right)$, $\frac{{\partial {\mathbf{\bar y}}\left( t \right)}}{{\partial {\theta _k}}} = {\text{vec}}\left( {{{{\mathbf{Y'}}}_{{\theta _k}}}\left( t \right)} \right)$, $\frac{{\partial {\mathbf{\bar y}}\left( t \right)}}{{\partial {r_k}}} = {\text{vec}}\left( {{{{\mathbf{Y'}}}_{{r_k}}}\left( t \right)} \right)$.
		\subsection{Calculate ${{\mathbf{F}}_B}\left( {\mathbf{\mu}} \right)$ and Jacobian matrix ${\mathbf{J}}$}
		First, the $\frac{{\partial {\mathbf{\tilde y}}\left( t \right)}}{{\partial {\bm{\mu}_B}}}$ in (\ref{eq36}) can be expressed as
		\begin{equation}\label{eq41}
			\begin{aligned}
				\frac{{\partial {\mathbf{\tilde y}}\left( t \right)}}{{\partial {\bm{\mu}_B}}} = \left[ {\begin{array}{*{20}{c}}
						{\frac{{\partial {\mathbf{\tilde y}}_{B,x}^{}\left( t \right)}}{{\partial {\theta _B}}}}&{{{\mathbf{0}}_{N \times 1}}} \\
						{{{\mathbf{0}}_{N \times 1}}}&{\frac{{\partial {\mathbf{\tilde y}}_{B,z}^{}\left( t \right)}}{{\partial {\varphi _B}}}}
				\end{array}} \right]
			\end{aligned}
		\end{equation}
		where ${{\mathbf{0}}_{N \times 1}} \in {\mathbb{R}^{{\text{N}} \times {\text{1}}}}$ is a zero vector. The partial derivatives of ${\mathbf{\tilde y}}_{B,x}^{}\left( t \right)$ and ${\mathbf{\tilde y}}_{B,z}^{}\left( t \right)$ with respect to ${\theta _B}$ and ${\varphi _B}$ are
		\begin{equation}\label{eq42}
			\begin{aligned}
				\frac{{\partial {\mathbf{\tilde y}}_{B,x}^{}\left( t \right)}}{{\partial {\theta _B}}} = \sqrt {1 - \delta } \frac{{\partial {{\mathbf{H}}_x}}}{{\partial {\theta _B}}}{{\mathbf{W}}_t}{\mathbf{\bar y}}\left( t \right)
			\end{aligned}
		\end{equation}
		\begin{equation}\label{eq43}
			\begin{aligned}
				\frac{{\partial {\mathbf{\tilde y}}_{B,z}^{}\left( t \right)}}{{\partial {\varphi _B}}} = \sqrt {1 - \delta } \frac{{\partial {{\mathbf{H}}_z}}}{{\partial {\varphi _B}}}{{\mathbf{W}}_t}{\mathbf{\bar y}}\left( t \right)
			\end{aligned}
		\end{equation}
		\begin{equation}\label{eq44}
			\scalebox{0.85}{$\displaystyle
				\begin{aligned}
					\frac{{\partial {{\mathbf{H}}_x}}}{{\partial {\theta _B}}} = {\gamma _x}{{\mathbf{b}}_x}\left( {{\varphi _B},{\theta _B}} \right) \odot \left( { - j\sin \left( {{\varphi _B}} \right)\sin \left( {{\theta _B}} \right){\mathbf{u}}} \right)*{\mathbf{b}}_R^{\text{H}}\left( {{\varphi _{R,B}},{\theta _{R,B}}} \right)
				\end{aligned}
				$}
		\end{equation}
		\begin{equation}\label{eq45}
			\begin{aligned}
				\frac{{\partial {{\mathbf{H}}_z}}}{{\partial {\varphi _B}}} = {\gamma _z}{{\mathbf{b}}_z}\left( {{\varphi _B}} \right) \odot \left( { - j\sin \left( {{\varphi _B}} \right){\mathbf{u}}} \right)*{\mathbf{b}}_R^{\text{H}}\left( {{\varphi _{R,B}},{\theta _{R,B}}} \right)
			\end{aligned}
		\end{equation}
		where ${\mathbf{u}} = {\left[ {\frac{{\left( {N - 1} \right)}}{2}, \cdots ,1, \cdots , - \frac{{\left( {N - 1} \right)}}{2}} \right]^{\text{T}}}$.
		
		After calculating the matrix ${{\mathbf{F}}_B}\left( {\bm{\mu}}_B \right)$, the Jacobian matrix ${\mathbf{J}}$ needs to be computed. To obtain each element in the matrix ${\mathbf{J}}$, it is necessary to express ${\theta _B}$ and ${\varphi _B}$ in terms of ${\mathbf{p}}_R^{}$, namely ${\theta _B} = \pi  - {\text{arctan2}}\left( {\left( {{{\left[ {{\mathbf{p}}_B^{}} \right]}_2} - {{\left[ {{\mathbf{p}}_R^{}} \right]}_2}} \right),\left( {{{\left[ {{\mathbf{p}}_B^{}} \right]}_1} - {{\left[ {{\mathbf{p}}_R^{}} \right]}_1}} \right)} \right)$, ${\varphi _B} = \pi  - {\text{arccos}}\left( {{{\left( {{{\left[ {{\mathbf{p}}_B^{}} \right]}_3} - {{\left[ {{\mathbf{p}}_R^{}} \right]}_3}} \right)} \mathord{\left/
					{\vphantom {{\left( {{{\left[ {{\mathbf{p}}_B^{}} \right]}_3} - {{\left[ {{\mathbf{p}}_R^{}} \right]}_3}} \right)} {\left\| {{\mathbf{p}}_R^{} - {\mathbf{p}}_B^{}} \right\|}}} \right.
					\kern-\nulldelimiterspace} {\left\| {{\mathbf{p}}_R^{} - {\mathbf{p}}_B^{}} \right\|}}} \right)$. The specific derivations of HRIS position parameters are as follows:
		\begin{equation}\label{eq46}
			\begin{aligned}
				\frac{{\partial {\theta _B}}}{{\partial {x_R}}} = \frac{{{{\left[ {{{\mathbf{p}}_B}} \right]}_2} - {{\left[ {{{\mathbf{p}}_R}} \right]}_2}}}{{\left\| {{{\left[ {{{\mathbf{p}}_R}} \right]}_{1:2}} - {{\left[ {{{\mathbf{p}}_B}} \right]}_{1:2}}} \right\|_{}^2}}
			\end{aligned}
		\end{equation}
		{\begin{equation}\label{eq47}
				\begin{aligned}
					\frac{{\partial {\theta _B}}}{{\partial {y_R}}} = \frac{{{{\left[ {{{\mathbf{p}}_R}} \right]}_1} - {{\left[ {{{\mathbf{p}}_B}} \right]}_1}}}{{\left\| {{{\left[ {{{\mathbf{p}}_R}} \right]}_{1:2}} - {{\left[ {{{\mathbf{p}}_B}} \right]}_{1:2}}} \right\|_{}^2}}
				\end{aligned}
		\end{equation}}
		\begin{equation}\label{eq48}
			\begin{aligned}
				\frac{{\partial {\theta _B}}}{{\partial {z_R}}} = 0
			\end{aligned}
		\end{equation}
		\begin{equation}\label{eq49}
			\begin{aligned}
				\frac{{\partial {\varphi _B}}}{{\partial {x_R}}} = \frac{{\left( {{{\left[ {{{\mathbf{p}}_R}} \right]}_3} - {{\left[ {{{\mathbf{p}}_B}} \right]}_3}} \right)\left( {{{\left[ {{{\mathbf{p}}_R}} \right]}_1} - {{\left[ {{{\mathbf{p}}_B}} \right]}_1}} \right)}}{{{{\left\| {{\mathbf{p}}_R^{} - {\mathbf{p}}_B^{}} \right\|}^2}{\left\| {{{\left[ {{{\mathbf{p}}_R}} \right]}_{1:2}} - {{\left[ {{{\mathbf{p}}_B}} \right]}_{1:2}}} \right\|} }}
			\end{aligned}
		\end{equation}
		\begin{equation}\label{eq50}
			\begin{aligned}
				\frac{{\partial {\varphi _B}}}{{\partial {y_R}}} = \frac{{\left( {{{\left[ {{{\mathbf{p}}_R}} \right]}_3} - {{\left[ {{{\mathbf{p}}_B}} \right]}_3}} \right)\left( {{{\left[ {{{\mathbf{p}}_R}} \right]}_2} - {{\left[ {{{\mathbf{p}}_B}} \right]}_2}} \right)}}{{{{\left\| {{\mathbf{p}}_R^{} - {\mathbf{p}}_B^{}} \right\|}^2}\, {\left\| {{{\left[ {{{\mathbf{p}}_R}} \right]}_{1:2}} - {{\left[ {{{\mathbf{p}}_B}} \right]}_{1:2}}} \right\|} }}
			\end{aligned}
		\end{equation}
		\begin{equation}\label{eq51}
			\begin{aligned}
				\frac{{\partial {\varphi _B}}}{{\partial {z_R}}} = \frac{{ - {\left\| {{{\left[ {{{\mathbf{p}}_R}} \right]}_{1:2}} - {{\left[ {{{\mathbf{p}}_B}} \right]}_{1:2}}} \right\|} }}{{{{\left\| {{\mathbf{p}}_R^{} - {\mathbf{p}}_B^{}} \right\|}^2}}}
			\end{aligned}
		\end{equation}

		\ifCLASSOPTIONcaptionsoff
		\newpage
		\fi

		\bibliographystyle{IEEEtran}
		\bibliography{ref}

\begin{thebibliography}{10}
\providecommand{\url}[1]{#1}
\csname url@samestyle\endcsname
\providecommand{\newblock}{\relax}
\providecommand{\bibinfo}[2]{#2}
\providecommand{\BIBentrySTDinterwordspacing}{\spaceskip=0pt\relax}
\providecommand{\BIBentryALTinterwordstretchfactor}{4}
\providecommand{\BIBentryALTinterwordspacing}{\spaceskip=\fontdimen2\font plus
\BIBentryALTinterwordstretchfactor\fontdimen3\font minus
  \fontdimen4\font\relax}
\providecommand{\BIBforeignlanguage}[2]{{%
\expandafter\ifx\csname l@#1\endcsname\relax
\typeout{** WARNING: IEEEtran.bst: No hyphenation pattern has been}%
\typeout{** loaded for the language `#1'. Using the pattern for}%
\typeout{** the default language instead.}%
\else
\language=\csname l@#1\endcsname
\fi
#2}}
\providecommand{\BIBdecl}{\relax}
\BIBdecl

\bibitem{2}
H.~J. Cho, Y.~Ahn, and B.~Shim, ``Transformer-aided mobile positioning for 6{G}
  ultra-dense networks,'' \emph{IEEE Trans. Veh. Technol.}, vol.~74, no.~4, pp.
  6839--6843, 2025.

\bibitem{3}
D.~Volgushev and G.~Fokin, ``Integrated communication, localization, sensing
  and synchronization in 6{G} cognitive wireless networks,'' in \emph{Proc.
  2024 Syste. Signal Synchronization, Generat. Process. Telecommun.
  (SYNCHROINFO)}, 2024, pp. 1--6.

\bibitem{Tuo2}
T.~Wu, C.~Pan, Y.~Pan, S.~Hong, H.~Ren, M.~Elkashlan, F.~Shu, and J.~Wang,
  ``Joint angle estimation error analysis and 3-{D} positioning algorithm
  design for mmwave positioning system,'' \emph{IEEE Internet Things J.},
  vol.~11, no.~2, pp. 2181--2197, 2024.

\bibitem{4}
X.~Guo, Q.~Shi, S.~Zhang, C.~Xing, and L.~Liu, ``User equipment assisted
  localization for 6{G} integrated sensing and communication,'' \emph{IEEE
  Trans. Commun.}, vol.~73, no.~11, pp. 12\,593--12\,607, 2025.

\bibitem{5}
Z.~Wu, Y.~Li, X.~Zhang, X.~Meng, X.~Lv, and Y.~Wu, ``Multiple anchors and
  {RIS}-aided localization method in complex {NLOS} environments,'' \emph{IEEE
  Internet Things J.}, vol.~11, no.~22, pp. 36\,922--36\,932, 2024.

\bibitem{105}
H.~Yan, H.~Chen, W.~Liu, S.~Yang, G.~Wang, and C.~Yuen, ``{RIS}-enabled joint
  near-field 3{D} localization and synchronization in {SISO} multipath
  environments,'' \emph{IEEE Trans. Green Commun. Netw.}, vol.~9, no.~1, pp.
  367--379, 2025.

\bibitem{6}
R.~Roy and T.~Kailath, ``{ESPRIT}-estimation of signal parameters via
  rotational invariance techniques,'' \emph{IEEE Trans. Acoust., Speech, Signal
  Process.}, vol.~37, no.~7, pp. 984--995, 1989.

\bibitem{7}
J.~Xie, H.~Tao, X.~Rao, and J.~Su, ``Efficient method of passive localization
  for near-field noncircular sources,'' \emph{IEEE Antennas Wireless Propag.
  Lett.}, vol.~14, pp. 1223--1226, 2015.

\bibitem{8}
R.~Schmidt, ``Multiple emitter location and signal parameter estimation,''
  \emph{IEEE Trans. Antennas Propag.}, vol.~34, no.~3, pp. 276--280, 1986.

\bibitem{9}
J.~He, M.~N.~S. Swamy, and M.~O. Ahmad, ``Efficient application of {MUSIC}
  algorithm under the coexistence of far-field and near-field sources,''
  \emph{IEEE Trans. Signal Process.}, vol.~60, no.~4, pp. 2066--2070, 2012.

\bibitem{10}
C.~Pan, H.~Ren, K.~Wang, J.~F. Kolb, M.~Elkashlan, M.~Chen, M.~Di~Renzo,
  Y.~Hao, J.~Wang, A.~L. Swindlehurst, X.~You, and L.~Hanzo, ``Reconfigurable
  intelligent surfaces for 6{G} systems: Principles, applications, and research
  directions,'' \emph{IEEE Commun. Mag.}, vol.~59, no.~6, pp. 14--20, 2021.

\bibitem{11}
A.~Umer, I.~Muursepp, M.~M. Alam, and H.~Wymeersch, ``Reconfigurable
  intelligent surfaces in 6{G} radio localization: A survey of recent
  developments, opportunities, and challenges,'' \emph{IEEE Commun. Surveys
  Tuts.}, vol.~27, no.~6, pp. 3526--3560, 2025.

\bibitem{12}
P.~Saikia, A.~Jee, K.~Singh, C.~Pan, W.-J. Huang, and T.~A. Tsiftsis,
  ``{RIS}-aided integrated sensing and communication systems: {STAR-RIS} versus
  passive {RIS}?'' \emph{IEEE Open J. Commun. Soc.}, vol.~5, pp. 7954--7973,
  2024.

\bibitem{Tuo3}
T.~Wu, C.~Pan, Y.~Pan, H.~Ren, M.~Elkashlan, and C.-X. Wang,
  ``Fingerprint-based mm{W}ave positioning system aided by reconfigurable
  intelligent surface,'' \emph{IEEE Wireless Commun. Lett.}, vol.~12, no.~8,
  pp. 1379--1383, 2023.

\bibitem{13}
Y.~Huang, J.~Yang, W.~Tang, C.-K. Wen, S.~Xia, and S.~Jin, ``Joint localization
  and environment sensing by harnessing {NLOS} components in {RIS}-aided mmwave
  communication systems,'' \emph{IEEE Trans. Wireless Commun.}, vol.~22,
  no.~12, pp. 8797--8813, 2023.

\bibitem{14}
M.~K. Ercan, A.~Pourafzal, M.~F. Keskin, S.~Gezici, and H.~Wymeersch,
  ``{RIS}-aided {NLOS} monostatic multi-target sensing under angle-doppler
  coupling,'' \emph{IEEE Trans. Veh. Technol.}, vol.~74, no.~12, pp.
  19\,141--19\,158, 2025.

\bibitem{Tuo1}
T.~Wu, C.~Pan, K.~Zhi, H.~Ren, M.~Elkashlan, C.-X. Wang, R.~Schober, and X.-H.
  You, ``Exploit high-dimensional {RIS} information to localization: What is
  the impact of faulty element?'' \emph{IEEE J. Sel. Areas Commun.}, vol.~42,
  no.~10, pp. 2803--2819, 2024.

\bibitem{106}
H.~Chen, T.~Gong, T.~Wu, M.~Elkashlan, B.~Liu, C.-B. Chae, K.-F. Tong, and
  K.-K. Wong, ``{FAS-ARIS}: Turning multipath challenges into localization
  opportunities,'' \emph{IEEE Trans. Netw. Sci. Eng.}, vol.~13, pp. 3756--3772,
  2026.

\bibitem{15}
J.~He, H.~Wymeersch, L.~Kong, O.~Silvén, and M.~Juntti, ``Large intelligent
  surface for positioning in millimeter wave {MIMO} systems,'' in \emph{Proc.
  2020 IEEE 91st Veh. Technol. Conf. (VTC2020-Spring)}, 2020, pp. 1--5.

\bibitem{16}
Y.~Jiang, W.~Yuan, and F.~Gao, ``{RIS} design for {CRB} optimization in source
  localization with electromagnetic interference,'' in \emph{Proc. 2023
  IEEE/CIC Int. Conf. Commun. China (ICCC)}, 2023, pp. 1--5.

\bibitem{17}
J.~Zhao, X.~Dong, J.~Qiu, J.~Luo, M.~Sun, Y.~Wang, and X.~Zhang, ``Sparse
  {RIS}-aided {DOA} estimation for {NLOS} scenario: a pseudo-inverse
  vectorization perspective,'' \emph{IEEE Trans. Veh. Technol.}, pp. 1--6,
  2025.

\bibitem{18}
F.~Wen, J.~Shi, G.~Gui, C.~Yuen, H.~Sari, and F.~Adachi, ``Joint {DOD} and
  {DOA} estimation for {NLOS} target using {IRS}-aided bistatic {MIMO} radar,''
  \emph{IEEE Trans. Veh. Technol.}, vol.~73, no.~10, pp. 15\,798--15\,802,
  2024.

\bibitem{19}
P.~Wang, J.~Fang, H.~Duan, and H.~Li, ``Compressed channel estimation for
  intelligent reflecting surface-assisted millimeter wave systems,'' \emph{IEEE
  Signal Process. Lett.}, vol.~27, pp. 905--909, 2020.

\bibitem{20}
Y.~You, Y.~Xue, L.~Zhang, X.~You, and C.~Zhang, ``Channel estimation for {RIS}
  assisted millimeter wave systems via {OMP} with optimization,'' \emph{IEEE
  Trans. Veh. Technol.}, vol.~72, no.~12, pp. 16\,783--16\,787, 2023.

\bibitem{21}
Z.~Tian, Z.~Zhang, and Y.~Wang, ``Low-complexity optimization for
  two-dimensional direction-of-arrival estimation via decoupled atomic norm
  minimization,'' in \emph{Proc. 2017 IEEE Int. Conf. Acoust., Speech, Signal
  Process. (ICASSP)}, 2017, pp. 3071--3075.

\bibitem{22}
Y.~Zheng, Q.~Wang, L.~Ren, Z.~Ma, and P.~Fan, ``{RIS} aided gridless {2D-DOA}
  estimation via decoupled atomic norm minimization,'' \emph{IEEE Trans. Veh.
  Technol.}, vol.~73, no.~10, pp. 15\,733--15\,738, 2024.

\bibitem{23}
X.~Wu, Y.~Liu, and H.~Zhang, ``Individual channel estimation in {RIS}-aided
  {MIMO} systems using atomic norm minimization,'' \emph{IEEE Trans. Commun.},
  vol.~73, no.~7, pp. 5111--5125, 2025.

\bibitem{24}
K.~T. Selvan and R.~Janaswamy, ``Fraunhofer and fresnel distances: Unified
  derivation for aperture antennas,'' \emph{IEEE Antennas. Propag. Mag.},
  vol.~59, no.~4, pp. 12--15, 2017.

\bibitem{25}
X.~Wei, L.~Dai, Y.~Zhao, G.~Yu, and X.~Duan, ``Codebook design and beam
  training for extremely large-scale {RIS}: Far-field or near-field?''
  \emph{China Commun.}, vol.~19, no.~6, pp. 193--204, 2022.

\bibitem{26}
H.~Zhang, N.~Shlezinger, F.~Guidi, D.~Dardari, and Y.~C. Eldar, ``6{G} wireless
  communications: From far-field beam steering to near-field beam focusing,''
  \emph{IEEE Commun. Mag.}, vol.~61, no.~4, pp. 72--77, 2023.

\bibitem{27}
C.~Ozturk, M.~F. Keskin, H.~Wymeersch, and S.~Gezici, ``{RIS}-aided near-field
  localization under phase-dependent amplitude variations,'' \emph{IEEE Trans.
  Wireless Commun.}, vol.~22, no.~8, pp. 5550--5566, 2023.

\bibitem{28}
M.~Luan, B.~Wang, Y.~Zhao, Z.~Feng, and F.~Hu, ``Phase design and near-field
  target localization for {RIS}-assisted regional localization system,''
  \emph{IEEE Trans. Veh. Technol.}, vol.~71, no.~2, pp. 1766--1777, 2022.

\bibitem{29}
J.~Kang, S.-W. Ko, and S.~Kim, ``Near-field localization with {RIS} via
  two-dimensional signal path classification,'' \emph{IEEE Trans. Wireless
  Commun.}, vol.~24, no.~4, pp. 3417--3432, 2025.

\bibitem{30}
H.~Zhang, N.~Shlezinger, I.~Alamzadeh, G.~C. Alexandropoulos, M.~F. Imani, and
  Y.~C. Eldar, ``Channel estimation with simultaneous reflecting and sensing
  reconfigurable intelligent metasurfaces,'' in \emph{Proc. 2021 IEEE 22nd Int.
  Workshop Signal Process. Advances Wireless Commun. (SPAWC)}, 2021, pp.
  536--540.

\bibitem{107}
R.~N. Challa and S.~Shamsunder, ``Passive near-field localization of multiple
  non-gaussian sources in {3-D} using cumulants,'' \emph{Signal Processing.},
  vol.~65, no.~1, pp. 39--53, 1998.

\bibitem{31}
Q.~Wu and R.~Zhang, ``Intelligent reflecting surface enhanced wireless network:
  Joint active and passive beamforming design,'' in \emph{Proc. 2018 IEEE
  Global Commun. Conf. (GLOBECOM)}, 2018, pp. 1--6.

\bibitem{32}
X.~Zhang and H.~Zhang, ``Hybrid reconfigurable intelligent surfaces-assisted
  near-field localization,'' \emph{IEEE Commun.Lett.}, vol.~27, no.~1, pp.
  135--139, 2023.

\bibitem{33}
G.~C. Alexandropoulos, I.~Vinieratou, and H.~Wymeersch, ``Localization via
  multiple reconfigurable intelligent surfaces equipped with single receive
  {RF} chains,'' \emph{IEEE Wireless Commun. Lett.}, vol.~11, no.~5, pp.
  1072--1076, 2022.

\end{thebibliography}

	\end{document}